\documentclass[floatfix,superscriptaddress,a4paper,
               nofootinbib,preprint]{revtex4-2}
\usepackage[colorlinks=true,linktocpage=true,linkcolor=blue,citecolor=blue,allcolors=blue]{hyperref}
\pdfoutput=1

\usepackage{graphicx}
\usepackage{amsmath}
\usepackage{amssymb}
\usepackage{bm}

\newcommand{\hsp}{\hspace*{1pt}}
\newcommand{\hspm}{\hspace*{.5pt}}
\newcommand{\ds}{\displaystyle}
\newcommand{\be}{\begin{equation}}
\newcommand{\ee}{\end{equation}}
\newcommand{\bel}[1]{\be\label{#1}}
\newcommand{\re}[1]{Eq.~(\ref{#1})}

\usepackage{color}

\begin{document}

\title
{
Bose-Einstein condensation in finite drops\\
of $\alpha$ particles
}

\author{L.~M. Satarov}
\affiliation{
Frankfurt Institute for Advanced Studies, D-60438 Frankfurt am Main, Germany}

\author{I.~N. Mishustin}
\affiliation{
Frankfurt Institute for Advanced Studies, D-60438 Frankfurt am Main, Germany}

\author{H. Stoecker}
\affiliation{
Frankfurt Institute for Advanced Studies, D-60438 Frankfurt am Main, Germany}
\affiliation{
Institut f\"ur Theoretische Physik,
Goethe Universit\"at Frankfurt, D-60438 Frankfurt am Main, Germany}
\affiliation{
GSI Helmholtzzentrum f\"ur Schwerionenforschung GmbH, D-64291 Darmstadt, Germany}

\begin{abstract}
Ground-state properties of finite drops of $\alpha$ particles (Q-balls) are studied
within a~field-theoretical approach in the mean-field approximation. The strong interaction of $\alpha$'s is described by the scalar field with
a sextic Skyrme-like potential. The radial profiles of scalar- and Coulomb fields
are found by solving the coupled system of Klein-Gordon and Poisson equations. The formation of shell-like nuclei, with vanishing density around the center, is predicted at high enough attractive strength of Skyrme potential. The equilibrium values of energy and baryon number of Q-balls and \mbox{Q-shells} are calculated for different sets of interaction parameters. Empirical binding energies of \mbox{$\alpha$-conjugate} nuclei are reproduced only if the gradient
term in the Lagrangian is strongly enhanced.
It~is demonstrated that this enhancement can be explained by a finite size of $\alpha$~particles.
\end{abstract}

\maketitle

\section{Introduction}
The formation of $\alpha$-particle {clusters} and quartic correlations in nuclei are
inte\-resting phenomena, that have been in the focus of theoretical and experimental studies for many years, see, e.g., Refs.~\cite{Fun09,Sog10,Ebr13}. {Several phenomenological and microscopic models have been suggested to describe clustering in nuclei and nuclear matter (see Ref.~\cite{Fre18}
for a~recent review). For example, a relativistic mean-field model with light clusters
has been used~\cite{Typ14} to calculate their radial profiles in heavy nuclei. It was conjectured
that such clusters occupy predominantly the dilute nuclear} {periphery.}

Especially interesting is the possibility of Bose-Einstein condensation (BEC) of {$\alpha$} clusters in so-called '$\alpha$-conjugate' nuclei, which have even proton and neutron numbers $Z=N=A/2$ (see Ref.~\cite{Oer10} and references therein). {Earlier we have studied} charge-neutral infinite systems of~$\alpha$~\cite{Sat17,Sat21} and $\alpha+N$~\cite{Mis16,Sat19,Sat20} particles {in the}~mean-field {approximation, using} Skyrme-like {effective} interactions.
{Both} the BEC and {the} liquid-gas phase transition {had been} considered.
{The} phase diagrams \mbox{of {the} $\alpha$\hspm -}~and $\alpha+N$ matter {were found to be} qualitatively similar to that observed for liquid~$^4$He\,.

{In the present paper} we consider finite systems of {charged} {bosons}, described by
a~scalar {mean} field {including} also Coulomb interactions. Such a field-theoretical approach was introduced originally in Refs.~\cite{Ros68,Fri76} for so-called Q-ball solitons. The ground states of Q-balls with attractive interactions of bosons were studied in more details by Coleman~\cite{Col85}. {Properties} of charged Q-balls, with Skyrme-like effective interactions, were considered in Ref.~\cite{Lee89}. Q-ball solutions were used to describe {heavy} $\alpha$-clustered nuclei in~Ref.~\cite{Mis18}.

One can consider Q-balls as localized, coherent superpositions of bosons similar to
atomic Bose condensates in magnetic traps. Properties of such condensates
are well described by the Gross-Pitaevskii equation~\cite{Gro61,Pit61}. As demonstrated in Ref.~\cite{Enq03},
the latter is a~non-relativistic {analogue} of the Klein-Gordon equation for scalar fields {of} \mbox{Q-balls}. {Introducing} Q-balls was rather successful in particle
physics and astrophysics (see, {e.g.,} Ref.~\cite{Nug20}). In~particular, gravitating Q-balls are {candidates} for bosonic stars~\cite{Lee92} and dark matter~\cite{Kus98}.
However, properties of charged Q-balls, and, especially, their stability criteria are {still}
not fully {explored}. For instance, a new type of soliton solutions, with vanishing
scalar density {around} the center ('Q-shells') have been found
recently~\cite{Aro09,Ish21,Hee21b} for certain {classes} of boson self-interactions.

In the present paper we {study} the
ground states (GS) of spherical Q-balls and \mbox{Q-shells} {made} of {charged} $\alpha$ particles.
{The} strong interactions of $\alpha$'s {are} described by a Skyrme-like {scalar} potential containing attrac\-tive and repulsive terms. First we use interaction parameters determined in Ref.~\cite{Sat21} by fitting the~microscopic calculations~\cite{Cla66} for homogeneous uncharged  $\alpha$ matter. Then, the sensitivity of {the} results to variation of these parameters {is analyzed}.

We emphasize that $\alpha$-clustered nuclei, {are} not necessa\-rily the stable {ground states}. {Some} of such nuclei {may} be metastable excited states, as{, e.g.,} the~$3\hspm \alpha$~Hoyle state in $^{12}$C {at} excitation energy 7.65~MeV. {Similarly,} $\alpha$-clustered excited states may exist in heavier nuclei, {even} with charge numbers {of} up to $Z\sim 80$\,, as follows from the analysis {in} Ref.~\cite{Oer10}\,. {It was shown within the density functional approach~\cite{Ebr14} that cluster formation probabilities indeed become larger in excited states of $\alpha$-conjugate nuclei.}
{As demonstrated in~Ref.~\cite{Mar16,Zai21},  $\alpha$-clustered} nuclei can be created in heavy-ion reactions at intermediate {and relativistic}
energies. {The} separation of $\alpha$-clustered and nucleonic states of~nuclear matter can be explained by the existence of {a} potential barrier between these two phases~\cite{Sat20}.

The paper is organized as follows: The model Lagrangian is introduced in {Sec.}~II\hspm A,
where {the} equations of motion for the scalar field and electrostatic potential {are also derived}. {In Sec.~II\hspm B we show} how the
particle numbers and binding energies of Q-balls are calculated.
{In Sec.~II\hspm C we discuss} a limiting case of homogeneous, uncharged $\alpha$~matter and introduce the constraints on the interaction parameters.

The radial profiles
of {the} baryon {density}, Coulomb {potential} and particle effective {mass} are calculated
in Sec.~III\hspm A and III\hspm B for {both}, Q-ball and Q-shell configurations.
{Their binding energies are analyzed in Sec.~IV\hspm A for different sets of model parameters.
Section~IV\hspm B~de\-mon\-strates} that empirical binding energies of~$\alpha$-conjugate nuclei can be reproduced {only} {when the gradient term in the effective Lagrangian is significantly enhanced.} It~is {argued} that such enhancement can be explained by {finite-size} effects in
the~$\alpha\alpha$~interaction. {Life times of metastable Q-balls are estimated in~Sec.~V.}

{The} numerical procedure is {explained} in Appendix A. The surface tension {coefficient} of cold $\alpha$~matter is calculated {ana\-lytically} and compared with {the} corresponding {value for} {isospin-symmetric} nuclear matter in Appendix B.

\section{Charged Q-balls in the mean-field approximation}
\label{sec-model}
\subsection{Equations of motion for scalar and Coulomb fields}

 In this paper {finite} systems of charged, massive bosonic particles
 {are studied by}
 \mbox{taking} into account {both} strong- and Coulomb interactions. All numerical calculations
 {are performed}  for finite systems of $\alpha$ particles.
  Below {we apply a field-theoretical approach denoting by} $\phi\hspm (x)$ and $A^\nu (x)$ the scalar and electromagnetic fields at {the} space-time point $x^\nu=(t,\bm{r})^\nu$. Generally, the Lagrangian density {of a bosonic} system can be written as~\cite{Lee89} ($\hbar=c=1$):
 \bel{lagd0}
{\cal L}=\frac{1}{2}\hspm D_\nu\phi \left(D^\nu\phi\right)^* -U(|\phi|)+\cal{L}_{\rm em}\,,
\ee
where $D_\nu=\partial_\nu -iqA_\nu$~($q=2e$ is the charge of {the} $\alpha$ particle), $U(|\phi|)$ is the mean-field potential, {which describes}
strong self-interactions of $\alpha$'s, and the last, electromagnetic, term reads
\bel{lagde}
{\cal L}_{\rm em}=-\frac{1}{16\pi}F_{\nu\sigma}\hspm F^{\nu\sigma},~~~F_{\nu\sigma}=\partial_\nu A_\sigma-\partial_\sigma A_\nu\,.
\ee
{Obviously}, the Lagrangian (\ref{lagd0}) is {locally} gauge-invariant. The corresponding conserved current is
\bel{curd}
J_\nu=\textrm{Im}\left(\phi^* D_\nu\phi\right)\,.
\ee

{In the following} only GS of spherical- and nonrotating Q-balls at zero temperature
{are studied}. In this case one can write
$\phi=e^{i\mu t}\varphi\hspm (r)$ and $A_\nu=A\hspm (r)\hspm\delta_{\nu,0}$, where $\varphi$ and $A$
are positive (real) func\-tions, and $\mu$ is the chemical potential\hsp\footnote
{
In the literature on Q-balls, it is usually denoted {by} $\omega$\hsp.
}. {Then} one has $J_\nu =n\hspm (r)\hspm\delta_{\nu,0}$, where $n=J_0$ is the number density of the $\alpha$ particles:
\bel{numd}
n=(\mu-q\hspm A)\hsp\varphi^2.
\ee
{The} Lagrangian density {can be written} as
\bel{lagd}
{\cal L}=\frac{1}{2}\hspm (\mu-q\hspm A)^2\varphi^2-\frac{1}{2}(\bm{\nabla}\varphi)^2-U(\varphi)+\frac{1}{8\pi}(\bm{\nabla}A)^2.
\ee
{Variation with respect to $\varphi$ and $A$ leads} to the following coupled equations:
\begin{eqnarray}
&&\Delta\hspm\varphi + (\mu-q\hspm A)^2\hsp\varphi=U^\prime (\varphi),\label{eom1}\\
&&\Delta A+4\pi\hspm q\hspm n=0\,.\label{eom2}
\end{eqnarray}
Equation~(\ref{eom1}) can be rewritten in the form of {the} Klein-Gordon equation (KGE),
with the effective mass squared
\bel{efm2}
M^2=U^\prime (\varphi)/\varphi\,.
\ee
Equation (\ref{eom2}) is the Poisson equation with the charge density $q\hspm n$, where $n$ is defined in~\re{numd}.
Stable Q-balls are characterized by the following boundary conditions, at small and large~$r$:
\begin{eqnarray}
&&\varphi^\prime\hspm (0)=0,~~A^\prime (0)=0,\\
&&\varphi\hsp (r),~A\hsp (r)\to 0~~\textrm{at}~r\to\infty\hspm .
\end{eqnarray}

Using Eqs. (\ref{eom1}) {and} (\ref{efm2}), one can see that {for $\Delta\varphi=0$} the KGE {is satisfied if} \mbox{$\mu-qA=M$}. Therefore, the homogeneous {solution} ($\varphi=\textrm{const}$) is possible only {when}
\bel{hbec}
\mu=M+qA\,.
\ee
This {relation may be regarded as} a generalization of the well-known BEC condition \mbox{$\mu=M$}~\cite{Sat21}, for {uncharged bosonic matter. Note that \re{hbec}} is gauge-invariant.
{Our} calculations show (see Fig.~\ref{fig2} {below}) that Eq.~(\ref{hbec}) holds approximately,  for central regions of large Q-balls\hspm .

\subsection{Particle number and binding energy of Q-balls}

The total number of particles in a Q-ball and its baryon number $B$ are found by integrating the density $n$ over the whole volume:
\bel{tnum}
Q=B/4=4\pi\hspace*{-0.5ex}\int\limits_0^\infty (\mu-qA)\hsp\varphi^2\hspm r^2\hspm dr.
\ee
Note, that in our grand canonical approach, $Q$ is, in general, {a} noninteger {quantity}. One should {bear} in mind that the mean-field
approximation, used in this paper, is not justified for small Q-balls, with~$Q\lesssim 1$\hsp .

{From} the Lagrangian, the energy-momentum tensor $T_{\nu\sigma}$ of Q-balls as well {as} profiles of {the} energy density $\varepsilon$ and pressure $p$ {can be calculated}. One gets (see, for details, Ref.~\cite{Log20})
\begin{eqnarray}
&&T_{00}=\varepsilon=\varepsilon_k+U+\varepsilon_{\rm gr}+\varepsilon_c\,,\label{epsp}\\
&&T_{rr}=\varepsilon_k-U+\varepsilon_{\rm gr}-\varepsilon_c=p+\frac{4}{3}\hsp(\varepsilon_{\rm gr}-\varepsilon_c)\,.\label{prep}
\end{eqnarray}
{Here}
\bel{endc}
\varepsilon_k=\frac{1}{2}\hsp (\mu-qA)^2\varphi^2,~~\varepsilon_{\rm gr}=\frac{1}{2}\hsp (\bm{\nabla}\varphi)^2,
~~\varepsilon_{\rm c}=\frac{1}{8\pi}\hsp (\bm{\nabla}A)^2
\ee
are, respectively, the kinetic, gradient and Coulomb contributions {to the} energy density.

The total energy $E$ and the binding energy per particle $W$ are obtained by integrating $\varepsilon$ over the whole volume:
\bel{tenq}
E=(m-W)\hsp Q=4\pi\int\limits_0^\infty \varepsilon\hspm r^2\hspm dr\hsp ,
\ee
{where} $m\simeq 3727.3~\textrm{MeV}$ is the mass of a single $\alpha$ particle.
The binding energy per {nucleon} equals $W_B=m_N-E/B=(W+B_\alpha)/4$, where $m_N\simeq 938.9~\textrm{MeV}$ is the nucleon mass and $B_\alpha=4\hspm m_N-m~\simeq 28.3~\textrm{MeV}$ is the binding energy of {the} $\alpha$ particle.

Using Eqs.~(\ref{eom1})--(\ref{eom2}), (\ref{tnum})--(\ref{tenq}), one can prove the validity of {the} thermodynamic relation~\cite{Col85}
\bel{ther}
dE=\mu\hsp d\hspm Q\hsp ,
\ee
{which connects} differentials of $E$ and $Q$ {at zero temperature}. This justifies our interpretation of~$\mu$ as the chemical potential.
The necessary conditions {for} the Q-ball stability can be written~as
\bel{csta}
\Delta\equiv m-\mu>0\,,~~~~W>0\,.
\ee
{The} first inequality implies that {the escape of} a single $\alpha$ particle from the Q-ball's surface to infinity is energetically forbidden~\cite{Lee89}\hspm\footnote
{
As discussed in Appendix~A, the region $\mu>m$ corresponds to continuum states, with~$\varphi$ oscillating at~$r\to\infty$\,.
However, these states are, in fact, metastable, due to presence of the Coulomb barrier.
}.
{The} calculations show that the conditions~(\ref{csta}) {are fulfilled} in the interval \mbox{$Q_{\rm min}<Q<Q_{\rm max}$} where $Q_{\rm min}$ and $Q_{\rm max}$
correspond, respectively, to~\mbox{$W=0$} and~$\Delta=0$\,.

Following Refs.~\cite{Sat21,Mis18}, the mean-field potential $U(\varphi)$ {is parameterized}
in the Skyrme-like form
\bel{skpt}
U(\varphi)=\frac{m^2}{2}\hsp\varphi^2-\frac{a}{4}\hsp\varphi^4+\frac{b}{6}\hsp\varphi^6,
\ee
 {where} $a$ and $b$ are positive parameters, {which determine}, respectively, the attractive and repulsive interactions of $\alpha$ particles.
Similar self-interactions of scalar bosons have been considered in Refs.~\cite{Lee89,Log20,Hee21a}.
{The above {expression} can be rewritten as}
\bel{skpt1}
\frac{\ds U}{\ds\varphi^2}=
\frac{b}{6}\left(\varphi^2-\varphi_0^2\right)^2+\frac{m^2}{2}\left(1-\frac{1}{\Lambda}\right).
\ee
{Here}
\bel{mpar}
\varphi_0=\sqrt{\frac{3\hspm a}{4\hspm b}}\,,~~~~~~\Lambda=\frac{16\hspm b\hspm m^2}{3\hspm a^2}
\ee
are the parameters which determine {the} main characteristics of {the} Q-balls, as well as properties of equilibrium, uncharged $\alpha$ matter (see {detailed calculations in} Ref.~\cite{Sat21}). In particular, $\varphi_0^2$ is the equilibrium scalar density
of a~homogeneous Bose-Einstein condensate at zero temperature.

Substitu\-ting~(\ref{skpt}) into \re{efm2} gives {the equation} for the effective mass squared
\bel{efm2a}
M^2=m^2-a\hsp\varphi^2+b\hsp\varphi^4.
\ee
One can see that $M^2(\varphi_0)=m^2\hsp (1-\Lambda^{-1})$ is nonnegative at $\Lambda\geqslant 1$\hspm\footnote
{
As pointed out in Ref.~\cite{Mis19}, the bosonic vacuum becomes unstable at $\Lambda<1$.
This does not occur for {the} Skyrme parameters expected {for} $\alpha$ interactions.
}.

\subsection{Properties of homogeneous, uncharged $\alpha$ matter}\label{infm}

{If the} Coulomb field is switched off ($q\to 0$), Q-balls could be arbitrary large. At zero temperature such hypothetical, 'uncharged' $\alpha$ matter has the GS characterized by
{a} spatially homogeneous scalar field, $\varphi=\varphi_0$. This {value} corresponds to the minimum of~$U(\varphi)/{\ds\varphi^2}$~\cite{Col85}. Indeed, applying formulas {of} preceding section for a~homogeneous, uncharged system with $\nabla\varphi=0, A=0$, one gets the {following} relations
\bel{inma}
\frac{\varepsilon}{n}=\dfrac{\mu}{2}+\dfrac{U(\varphi)}{\mu\hspm\varphi^2}\,,
~~~~~~~\mu=M(\varphi)\,.
\ee
Therefore, the minimum of the energy per particle $\varepsilon/n$ is {found} by mini\-mizing~$U/\varphi^2$.
The second condition in~\re{inma} is obtained from the KGE for homogeneous uncharged matter (see~Eqs.~(\ref{eom1}) and (\ref{efm2}))\hspm .
It is interesting that {it} coincides with the generalized condition of Bose-Einstein condensation {as} {introduced in} Ref.~\cite{Sat21}. By using Eqs.~(\ref{inma})
one obtains the following characteristics of the $\alpha$-matter GS
\bel{inma1}
\mu=\textrm{min}\left(\frac{\varepsilon}{n}\right)=m\sqrt{1-\frac{1}{\Lambda}}\,.
\ee

Adjusting the binding energy $W=m-\mu$ and {the} density $n=\mu\hsp\varphi^2_0$ to the GS properties of~homogeneous uncharged $\alpha$ matter, {as} obtained by microscopic
calculations in Ref.~\cite{Cla66}, {we} extracted the values
(for details, see Ref.~\cite{Sat21})
\bel{set1}
a=7853,~~~~~~b=78.94~\textrm{MeV}^{-2}\,.
\ee
Below this parameter choice {is denoted} as Set~I.

There {exists an} additional constraint {on} $\Lambda$~\cite{Sat17}, {which follows from the} obvious condition, {namely,} that $\alpha$ matter {has to} be less bound than
{isospin}-symmetric nuclear matter {in the GS}. It is commonly accepted that
{such matter} has {the} binding energy per baryon $W_{\rm SM}\simeq 16~\textrm{MeV}$. On the other hand, in accordance with~\re{inma1}, in the limit of large~$\Lambda$ the binding energy of $\alpha$ matter per particle equals $W\simeq m/(2\Lambda)$\,.
Therefore, one gets the condition
\bel{wbcn}
W_B\simeq\frac{1}{4}\left(\frac{m}{2\Lambda}+B_{\alpha}\right)<W_{\rm SM}\,.
\ee
This is equivalent to $\Lambda>52$ (approximately) or
\bel{cons1}
\frac{a}{\sqrt{b}}<1190~\textrm{MeV}\,.
\ee
{For} \mbox{$b\sim 80~\textrm{MeV}^{-2}$} (which is {close} to the value assumed in Set I) one obtains the con\-straint\hspm\footnote
{
{Note} that the parameters
$b=30.73~\textrm{MeV}^{-2}$ and \mbox{$a=3\cdot 10^4$} used in Ref.~\cite{Mis18} violate
the con\-straint~(\ref{cons1}), which results in {an} {unrealistically} large binding energy of~$\alpha$ matter $W_B\simeq 214~\textrm{MeV}$.
}~\mbox{$a\lesssim 1.1\cdot 10^4$}.


\section{Ground-state properties of charged Q-balls.}

\subsection{Radial profiles of density and effective mass}
\label{sec-bosm}

{Using the effective potential (\ref{skpt}) we have solved numerically coupled equations (\ref{eom1}) and~(\ref{eom2}) for different values of interaction parameters $a$ and $b$\hspm . Similar calculations} have been done earlier in Refs.~\cite{Lee89,Log20,Hee21a}.
It was shown that, in general, for a~given $\mu$ there {exist} two stable Q-ball configurations
with different effective radii, energies and particle numbers.
However, {the previous} authors did not analyze in detail the sensitivity to the {model} parameters and did not specify bosonic particles bound in Q-balls\,.
We also make comparison
with empirical data~\cite{Oer10,Oer06} on $\alpha$-conjugate nuclei.
\begin{table}[htb!]
\caption
{\label{tab1}Characteristics of Q-balls for different values of chemical potential {$\mu=m-\Delta$} {calculated for parameter Set I}\hspm .}
\vspace*{3mm}
\begin{tabular}{|c|c|c|c|c|}
\hline
~$\Delta\hspm,\,\textrm{MeV}$~&~$Q$~&~$W,\,\textrm{MeV}$~&~$\varphi\hspm (0)/\varphi_0$~&
~$qA\hspm (0)\hspm,\,\textrm{MeV}$~\\
\hline
\hspace*{1ex}$0$\hsp\protect\footnote
{
~The results for $\Delta=0$ {are obtained by setting} $\Delta=10^{-4}~\textrm{MeV}$.
}
&$~14.1~$&$~~~5.6~$&$~0.896~$&$~25.5~$\\[-2mm]
$2$    &$~11.8~$&$~~~6.6~$&$~0.924~$&$~22.4~$\\[-2mm]
$4$    & $~9.0~$&$~~~7.6~$&$~0.949~$&$~19.2~$\\[-2mm]
$6$    & $~6.7~$&$~~~8.6~$&$~0.971~$&$~15.9~$\\[-2mm]
$8$    & $~4.5~$&$~~~9.3~$&$~0.994~$&$~12.3~$\\[-2mm]
$10$   & $~2.4~$&$~~~9.6~$&$~1.022~$&$~8.0~$\\[-2mm]
$10.7$& $~1.2~$&$~~~8.8~$&$~1.042~$&$~5.0~$\\[-2mm]
$10.1$& $~0.52~$&$~~~6.5~$&$~1.057~$&$~2.8~$\\[-2mm]
$9.1$ & $~0.32~$&$~~~4.4~$&$~1.054~$&$~1.9~$\\[-2mm]
$5.5$ & $~0.14~$&$~~~0.01$&$~0.958~$&$~0.9~$\\
\hline
\end{tabular}
\end{table}
Characteristics of stable Q-balls, calculated for the parameter Set I~(see~\re{set1}) are shown in Table~\ref{tab1}\hspm .
{The} last two columns give {the} central values of {the} scalar- and Coulomb fields. {Note}
that $\varphi\hspm (0)\simeq\varphi_0$ {where $\varphi_0$ is} given by~\re{mpar} which implies a {rather} good accuracy of the 'thin wall' approximation~\cite{Lee89}, even for~\mbox{$Q\sim 1$}. According to~\mbox{Table I},
{the} maximum particle number for stable states $Q_{\rm max}\simeq 14$ corresponds to $\Delta=0$
{(i.e.~\mbox{$\mu=m$})}\,. {\mbox{However}, due}
to the presence of {the} Coulomb barrier {we expect} {the} formation of metastable Q-balls even for $\Delta<0$ (see {Sec.~\ref{msqb1}}).

\begin{figure}[htb!]
\centering
\includegraphics[trim=2cm 7.5cm 3cm 8cm,width=0.48\textwidth]{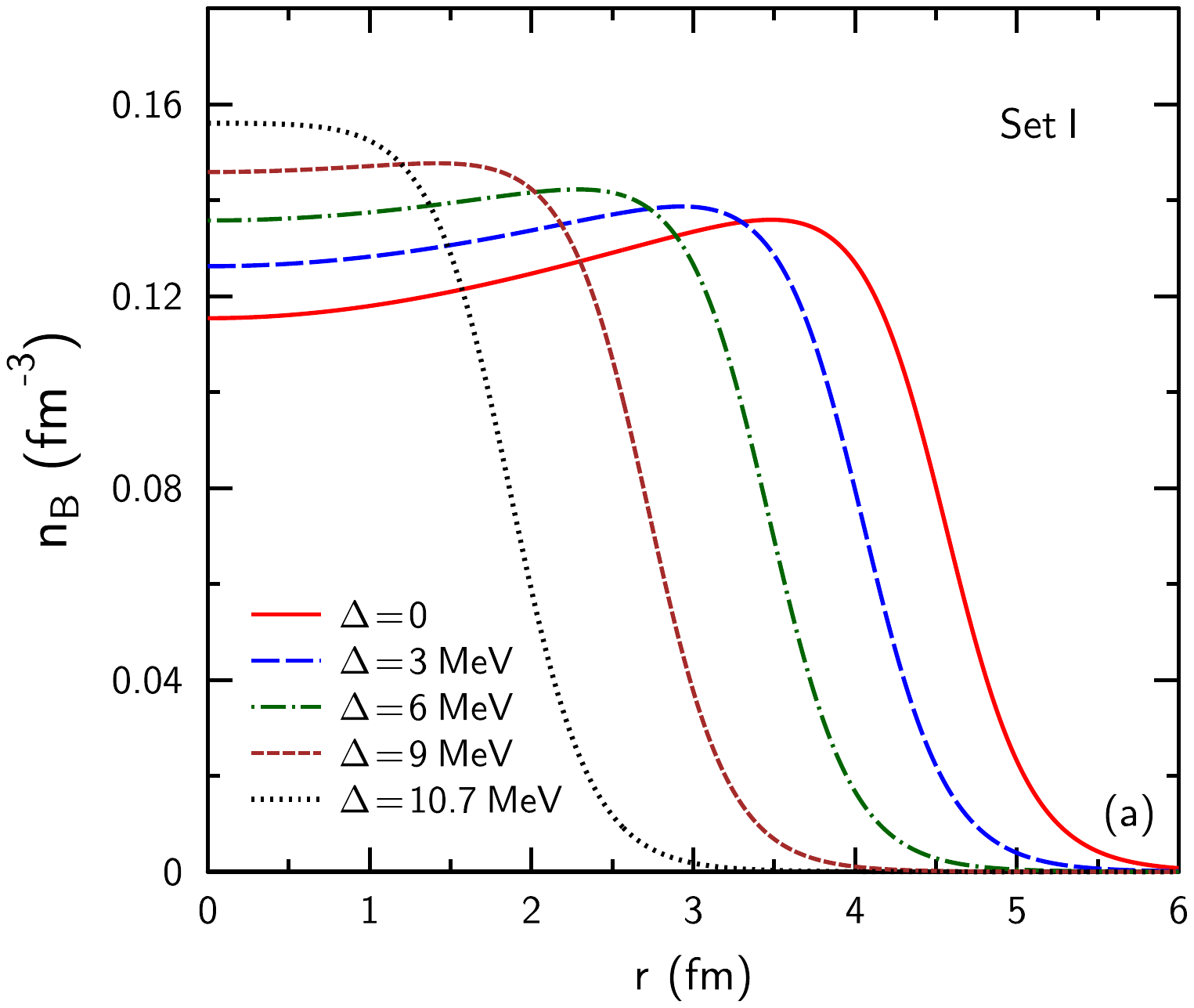}
\includegraphics[trim=2cm 7.5cm 3cm 8cm,width=0.48\textwidth]{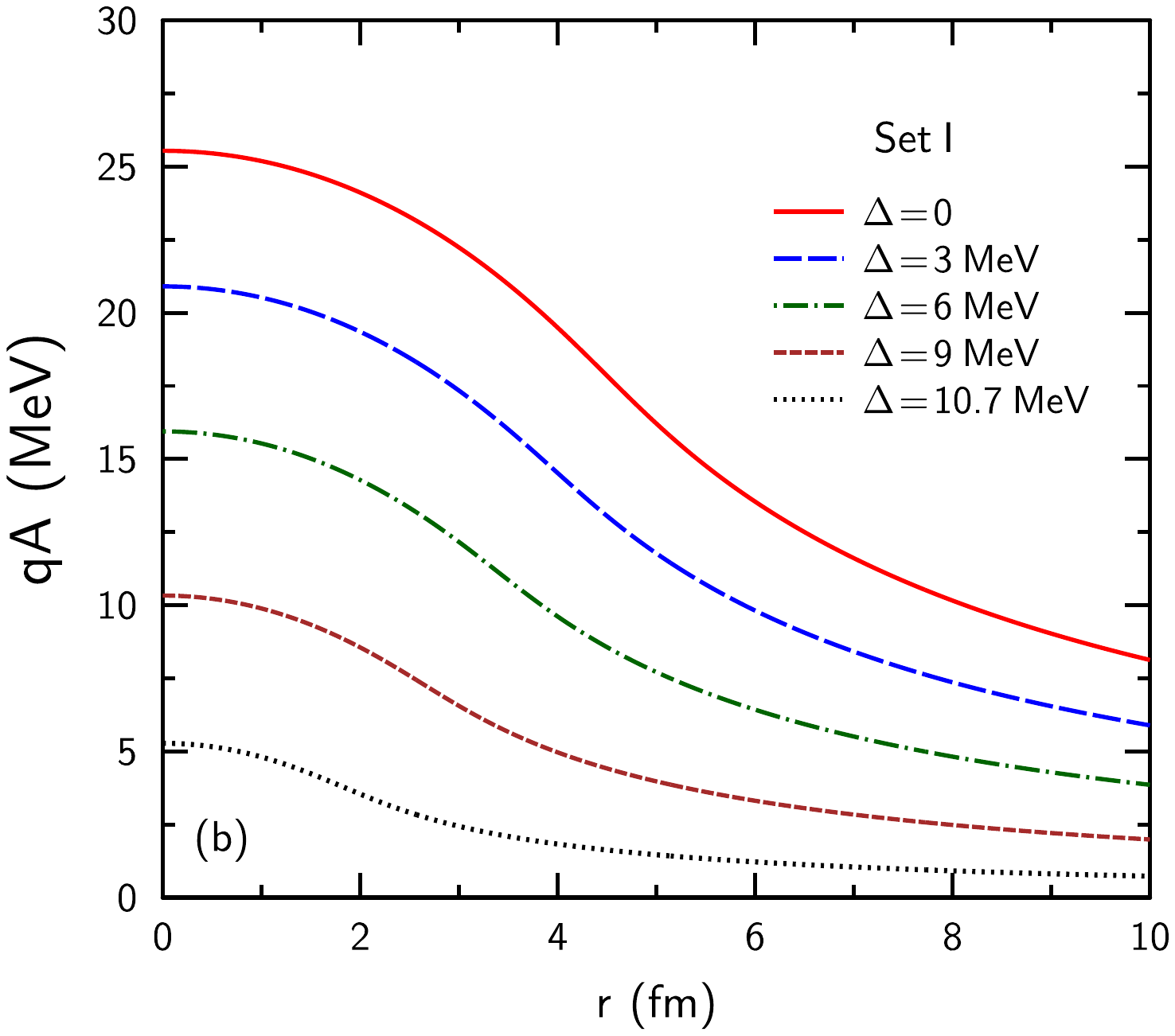}
\caption{
Radial profiles of the baryon density $n_B=4\hspm n$ (a) and Coulomb potential $qA$ (b) for different values of {the} chemical potential {$\mu=m-\Delta$}, {calculated {for}} the parameter Set I\hspm .
}\label{fig1}
\end{figure}

\begin{figure}[htb!]
\centering
\includegraphics[trim=2cm 7.5cm 3cm 8cm,width=0.6\textwidth]{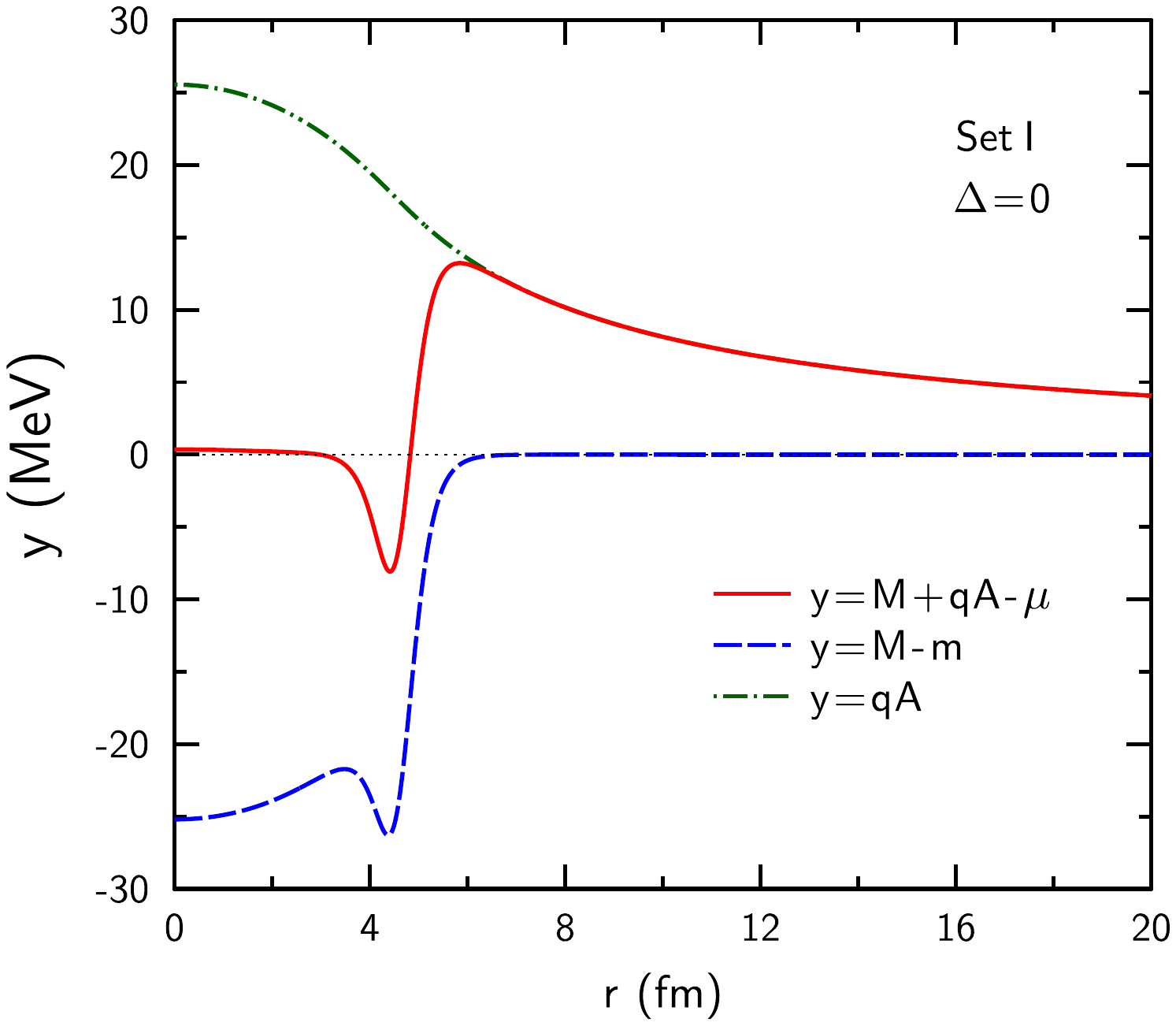}
\caption{
The {radial} profiles of $M+qA-\mu$ (the solid line), $M-m$ (the dashed curve)
{and $qA$ (the dash-dotted line)},
for $\Delta=m-\mu=0$ and the parameter Set I.
}\label{fig2}
\end{figure}
Figures~\ref{fig1} {and} \ref{fig2} show the profiles of the baryon density, Coulomb potential
and effective mass for different $\mu$. {One can see a central density suppression,
especially {well} visible for $\Delta\lesssim 5~\textrm{MeV}$. This effect is {due} to
the {repulsive} Coulomb interaction, {and it} does not appear for uncharged Q-balls (see,
e.g., Ref.~\cite{Lee89})\hspm . {A similar central depletion} of proton density was predicted in {the relativistic mean-field model of} Ref.~\cite{Ben99} for superheavy nuclei with~\mbox{$Z\sim 120$}. {According to Fig.~\ref{fig2}, in the case $\mu=m$, the Q-ball{'s} chemical potential~is, in fact, the result of cancellation between the attractive well of the depth $m-M$  and the repulsive Coulomb energy $qA$ at~$r<R\simeq 5~\textrm{fm}$\hspm\footnote
{
Note that {for} $\Delta=0$ the central electrostatic potential $qA(0)$ is close to the value \mbox{$3q^2Q/(2R)\simeq 24~\textrm{MeV}$
}
for a~homo\-geneously charged sphere with the radius $R$ and total charge~$Q$\,.  One can also see that the height of the Coulomb barrier at $r=R$ is approximately equal to $q^2Q/R\simeq 16~\textrm{MeV}$\hspm .
\label{elst1}}.
This cancellation does not occur at larger radii, where {the~nuclear potential disappears and} the Coulomb barrier dominates.}

\subsection{Two types of solutions: Q-balls and Q-shells}

We have {investigated possible solutions of field equations for different sets of
model parameters.}
It was found that a new class of {solutions} appears {for} large enough {values of the} attraction coefficient $a$. Namely, the model predicts {the} formation of {hollow,} shell-like structures with vanishing baryon density in the central {region}. To distinguish these two types of solutions, we call them Q-balls and Q-shells, respectively. {The} appearance of Q-shells was demonstrated in Refs.~\cite{Aro09,Ish21,Hee21b}.

\begin{figure}[htb!]
\centering
\includegraphics[trim=2cm 8cm 3cm 8cm,width=0.48\textwidth]{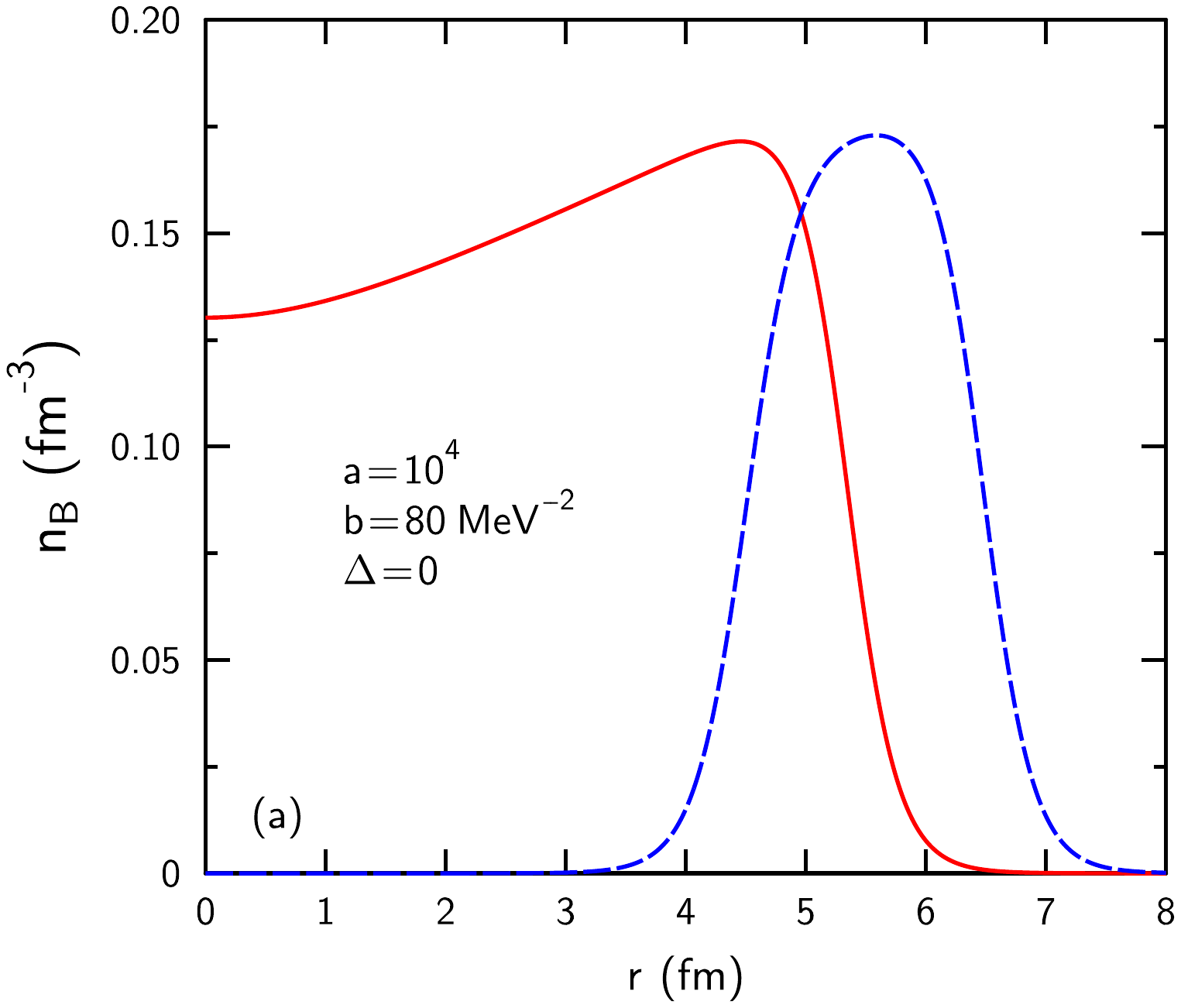}
\includegraphics[trim=2cm 8cm 3cm 8cm,width=0.48\textwidth]{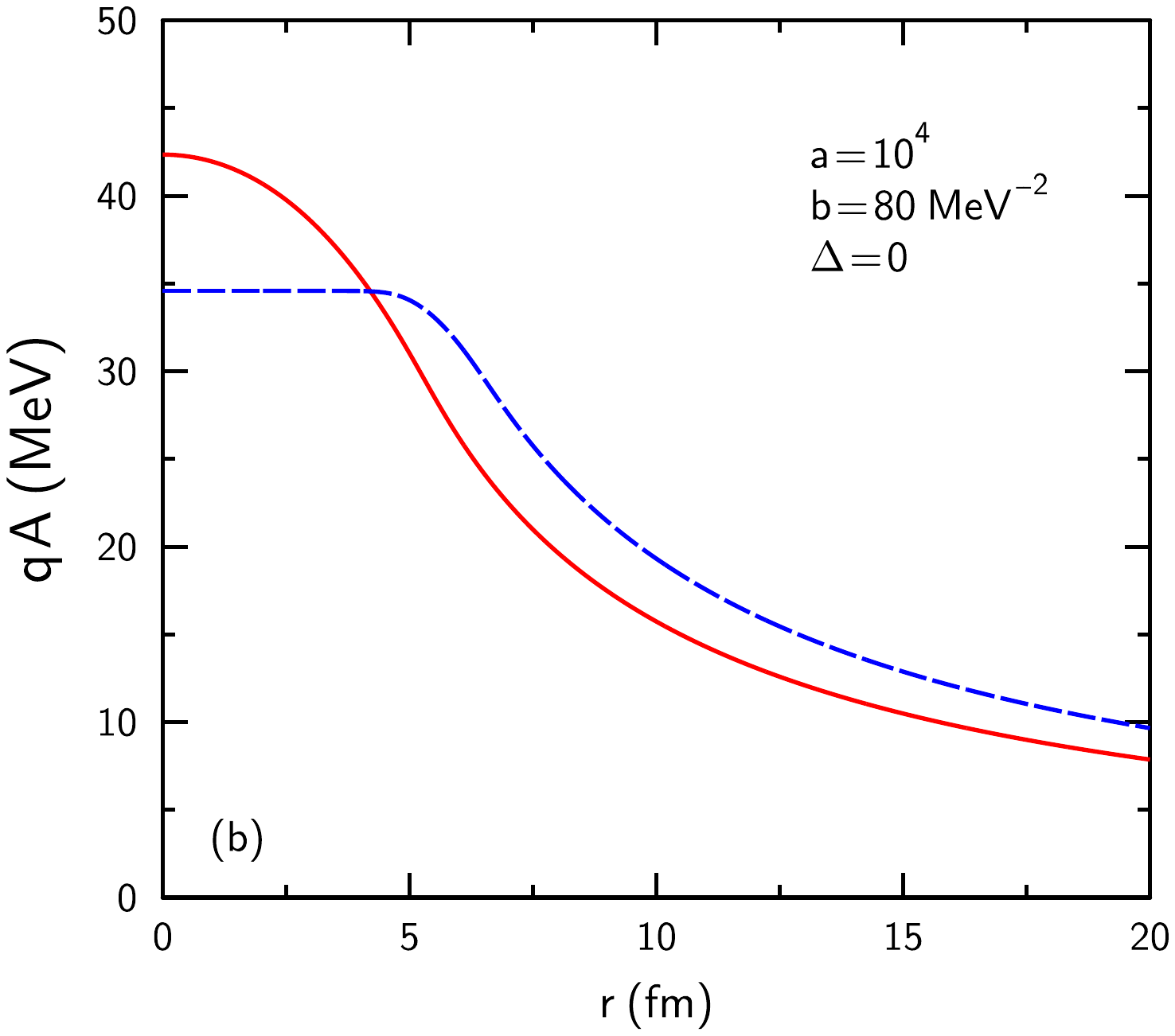}
\caption{
The profiles of baryon density (a) and Coulomb potential (b) for \mbox{$\mu=m$} and the para\-meters \mbox{$a=10^4$}, \mbox{$b=80~\textrm{MeV}^{-2}$}.
Solid and dashed lines correspond to the Q-ball and Q-shell solutions, respectively.
}\label{fig3}
\end{figure}

The calculations show that {stable} Q-shells do not exist for the parameter Set~I.
According to our analysis, at $b\simeq 80~\textrm{MeV}^{-2}$ they appear only for $a\gtrsim 9\cdot 10^3$ and small {enough} $\Delta$~values\,. Figure~\ref{fig3} {represents the} radial profiles of the baryon density and Coulomb potential, {as} obtained for \mbox{$a=10^4$}, \mbox{$b=80~\textrm{MeV}^{-2}$} and $\Delta=0$\hspm .
{These} parameters {yield} the values $Q\simeq 27.3$, \mbox{$W\simeq 10.1~\textrm{MeV}$} for the Q-ball solution, {while}  $Q\simeq 33.5, W\simeq 8.5~\textrm{MeV}$ for the Q-shell. {The}
{new shell-like} solution corresponds to larger (smaller) values of gradient (Coulomb) energy per particle.
\begin{figure}[htb!]
\centering
\includegraphics[trim=2cm 8cm 3cm 8cm,width=0.48\textwidth]{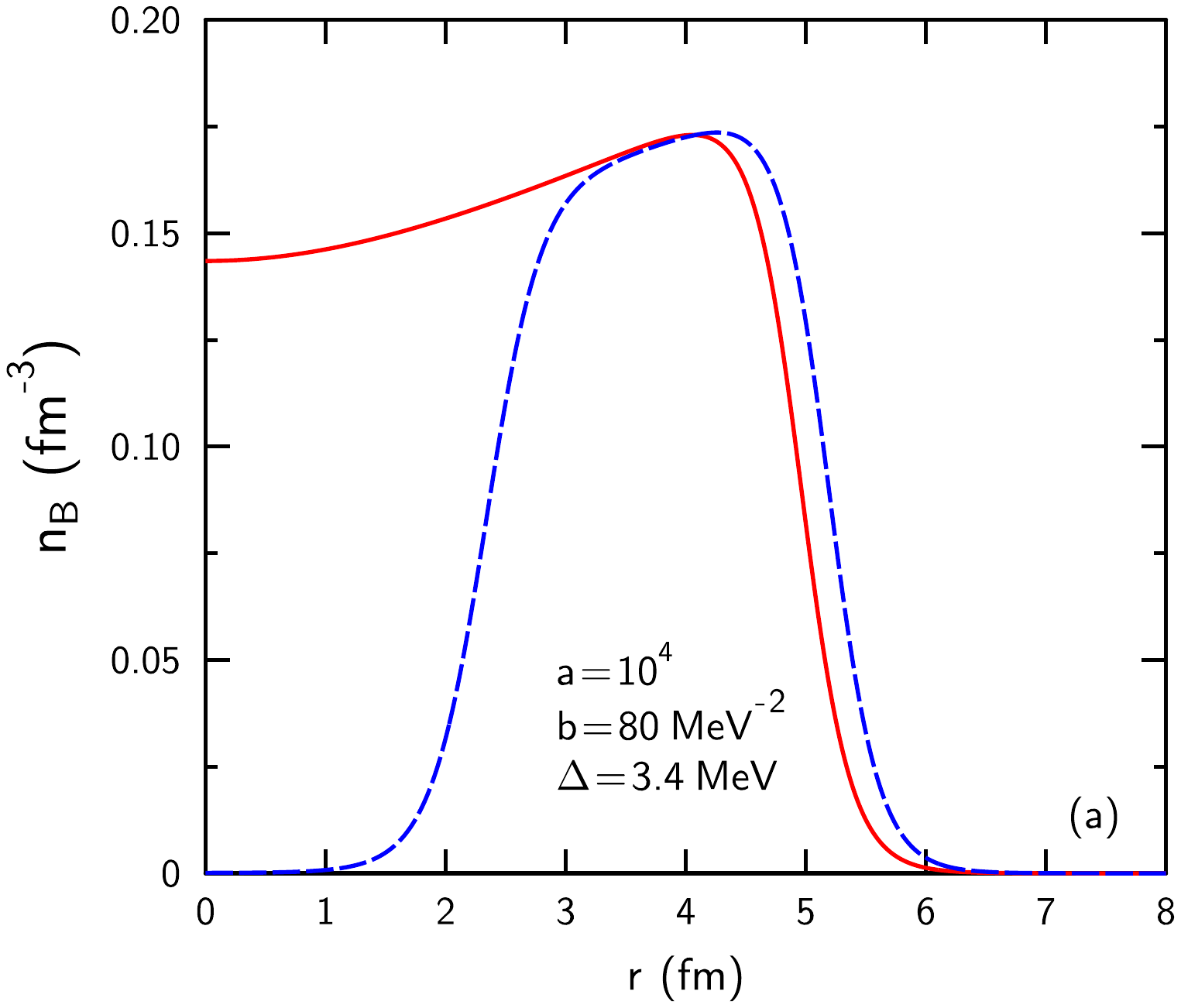}
\includegraphics[trim=2cm 8cm 3cm 8cm,width=0.48\textwidth]{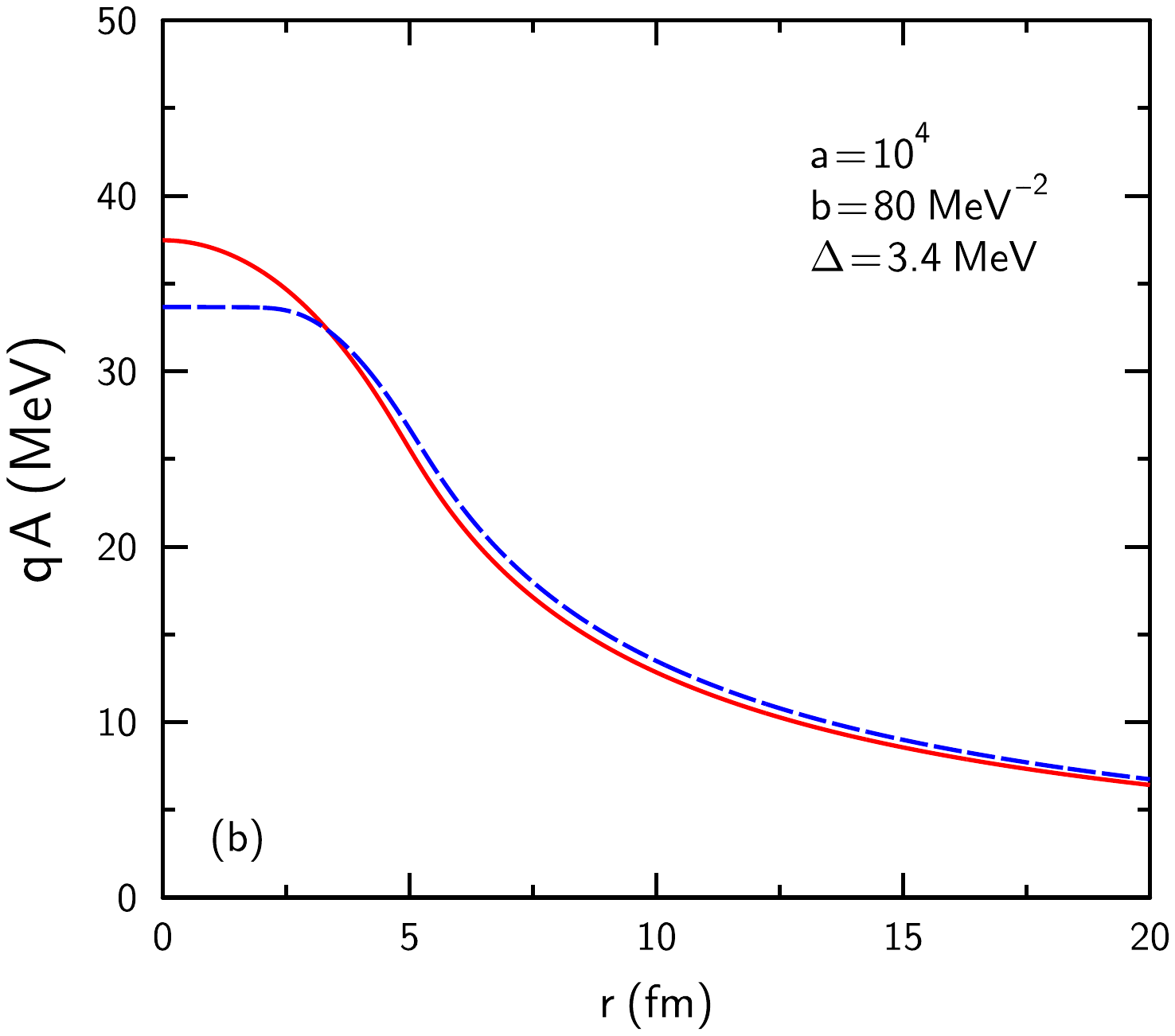}
\caption{
Same as Fig.~\ref{fig3}, but for $\Delta=3.4~\textrm{MeV}$\,.
}\label{fig4}
\end{figure}

{Figure~\ref{fig4}} shows the results for} the same parameters $a,b$, but {for} {nonzero} $\Delta$ {value, corresponding to smaller~$\mu$}.
{Reducing the} chemical potential gives rise to shifting the outer surface of~a~\mbox{Q-shell} {towards} the Q-ball's surface. This is seen in Fig.~\ref{fig4} {where} \mbox{Q-ball}
and Q-shell profiles {are considered} {for}~\mbox{$\Delta=3.4~\textrm{MeV}$}\hspm\footnote
{
At larger values of $\Delta$, Q-shell solutions disappear for the {values of} interaction parameters considered here.
}.
In this case {the model predicts} closer characteristics: $Q\simeq 23.4, W\simeq 11.5~\textrm{MeV}$ and \mbox{$Q\simeq 22.3, W\simeq 11.0~\textrm{MeV}$}, for the~\mbox{Q-ball} and Q-shell configurations, respectively.
These two configurations with close values of~$Q$ may be regarded as density isomers, separated by a~potential barrier.

{The sensitivity of Q-shell properties to the model parameters will be considered in the next section.}

\section{Systematics of binding energies of Q-balls and Q-shells}

\subsection{Comparison with empirical binding energies of $\alpha$-conjugate nuclei}

In this section we consider how the present model is able to reproduce the 'experimental'
data.  The GS binding energies of $\alpha$-conjugate nuclei (with \mbox{$Z=N=2\hspm Q$}) were compiled by von Oertzen {in Ref.}~\cite{Oer06}. It was conjectured that such nuclei can be regarded as made of~$\alpha$~particles. The compiled data are shown below
by dots.
\begin{figure}[htb!]
\centering
\includegraphics[trim=2cm 7.5cm 3cm 8cm,width=0.6\textwidth]{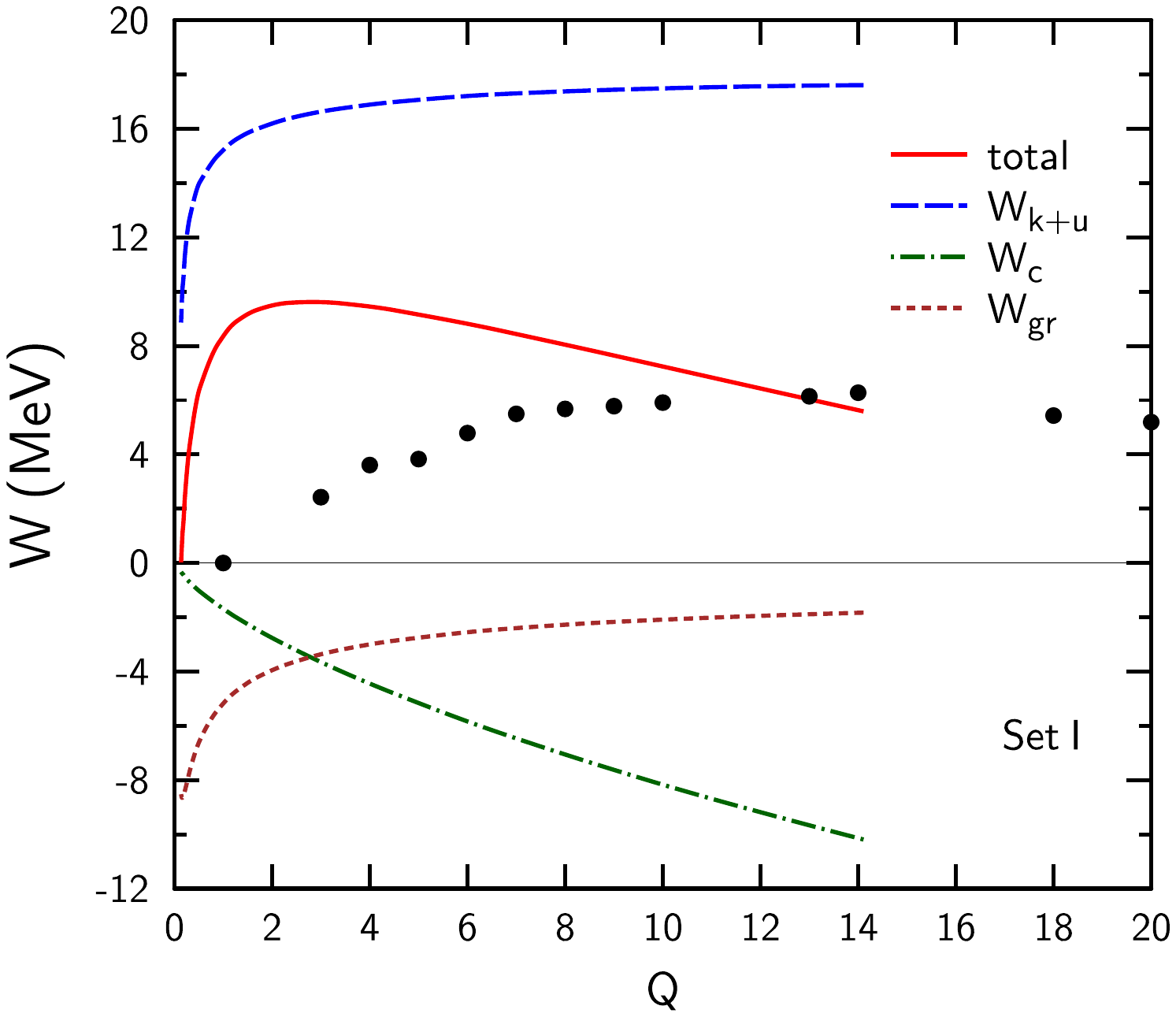}
\caption{
Binding energy per particle, $W$, as a function of {the} total number of $\alpha$'s, $Q$,
for parameter Set~I (the solid line). {The} combined contribution of the kinetic and potential
parts of {the Q-ball's} energy is shown by the long-dashed curve. The Coulomb and gradient
energies are shown by the dash-dotted and short-dashed lines, respectively.
Dots are empirical values of {the} binding energies of $\alpha$-conjugate nuclei with $A=4\hspm Q$ from Ref.~\cite{Oer06}\hsp .
}\label{fig5}
\end{figure}

Figure~\ref{fig5} {shows} the Q-ball binding energy $W$ as a function of $Q$
for the parameter Set I. {The} right end point of the $W(Q)$ curve, corresponding to the chemical
potential $\mu=m$, gives the~maxi\-mal {possible} {number} of $\alpha$ particles in a stable Q-ball. {At $\mu>m$,
Q-balls become metastable, {as} it is energetically possible {for} $\alpha$'s {to leave the system and go to} infinity. However, {at not too large $\mu-m$}, this process
is {strongly} suppressed by the Coulomb barrier (see Fig.~\ref{fig2})\hspm . {The} life times of these metastable
states {can} be very long as known from $\alpha$-decays of ordinary nuclei, see {Sec.~\ref{msqb1}}. Therefore, {the region~$\mu>m$ {is also included} in} our analysis.

{In Fig.~\ref{fig5} we show} separately different contributions to the \mbox{Q-ball's}
energy per particle. They were calculated by using the corresponding energy density
terms in~\re{endc}. {Note,} that the absolute value of the gradient energy {per particle} decreases, {while} the Coulomb contribution increases with $Q$. Obviously, our
mean-field approach is not reliable at $Q\lesssim 1$\hspm .

One can see, that the model predictions with the parameter Set I {strongly deviate from} the empirical data. In particular, binding energies of {lighter} nuclei, with $Q\lesssim {10}$,  are significantly overestimated.
{In principle, one can find such model parameters $a$ and $b$ which allow a~good
fit for any individual \mbox{$\alpha$-{conjugate}} nucleus, as, e.g., $^{52}$Fe ($Q=13$)
in Fig.~\ref{fig5}, but {then it is not possible to}
fit the empirical data {for other} $\alpha$-conjugate nuclei in the whole mass-number interval.
Nevertheless, the shape of the binding energy curve is qualitatively similar to that predicted {by}  Weizs\"acker's formula~\cite{Wei35} for ordinary nuclei. Namely, binding energies drop at
small~$Q$ due to the surface {energy}, and at large~$Q$~because of the Coulomb repulsion.

{Figure~\ref{fig6} demonstrates} the sensitivity of binding energies $W(Q)$ to the choice of {the} {interaction} parameters $a$ and~$b$. Again one can see that the model predictions are in strong disagreement with empirical data, and it is not possible to improve the fit  {just} by choosing parameters significantly different from {Set} I.

{By analyzing these results one can make the following conclusions:
first, the boundaries of the \mbox{Q-ball} and Q-shell stability ($\Delta=0$) shift to larger \mbox{Q-values} with increasing $a$ or decreasing $b$}\hsp\footnote
{
{As shown in Appendix~B this} {leads to increased} {values of the surface tension.}
}.
{Second, Q-shells appear only at large enough~$Q$, and  threshold Q-values
where they become more bound as compared to Q-balls
(see crosses in Fig.~\ref{fig6}) increase (decrease) with $a~(b)$\hspm .
Note, that in the case of the parameter Set I this threshold shifts to the region
of the Q-ball metastability, i.e, outside the end points of the dash-dotted
lines in Fig.~\ref{fig6}\hspm .}

Inspecting Figs.~\ref{fig5} and \ref{fig6} suggests that the large deviation from
the data is caused by too small surface energy, which is generated by the gradient term~(GT) $\varepsilon_{\rm gr}$ {(see~\re{endc})}.
Indeed, as shown in the next section,
agreement with {the} empirical data can be achieved by~a~signi\-ficant enhancement of the~GT.
\begin{figure}[htb!]
\centering
\includegraphics[trim=2cm 7.5cm 3cm 8cm,width=0.48\textwidth]{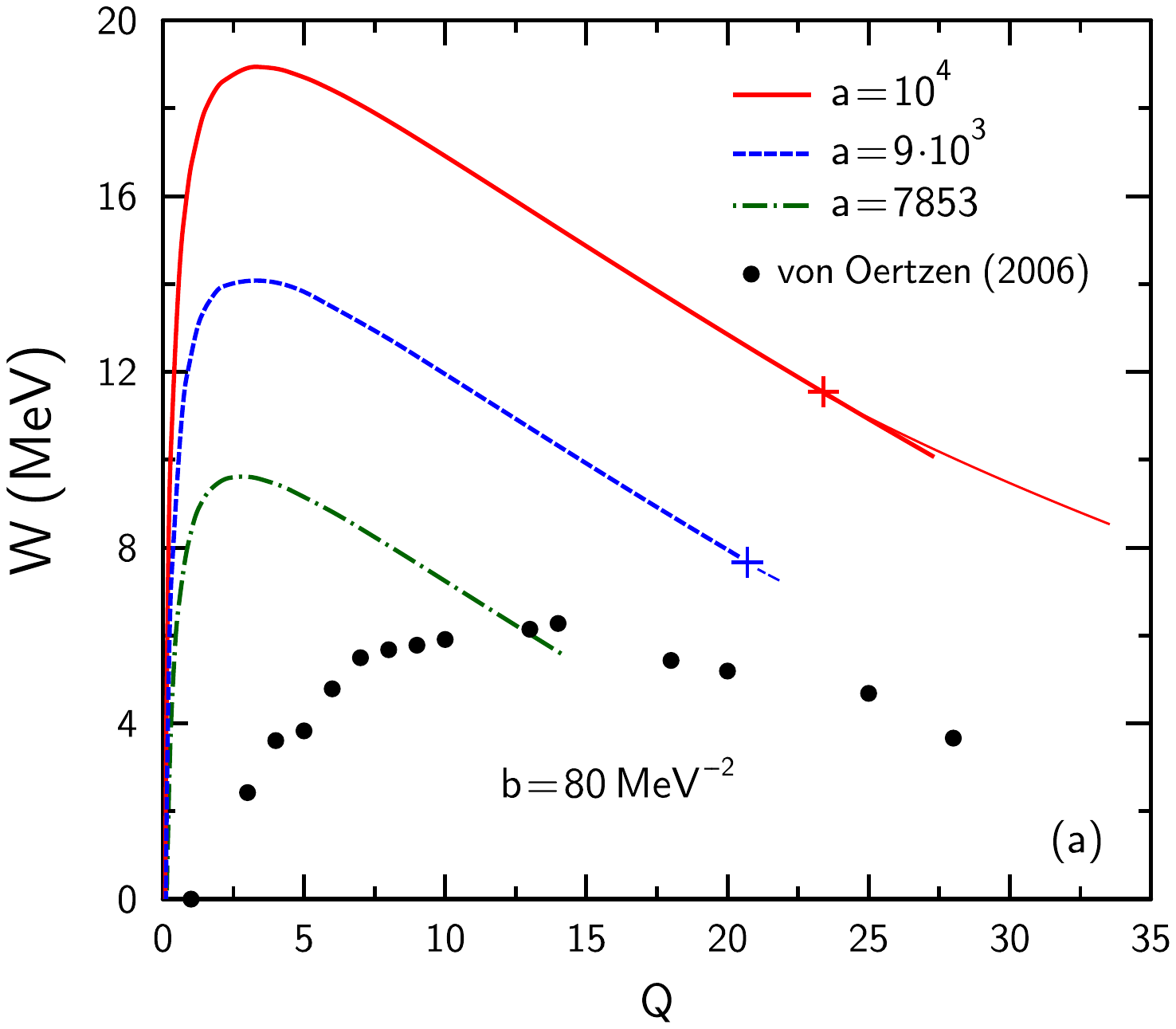}
\includegraphics[trim=2cm 7.5cm 3cm 8cm,width=0.48\textwidth]{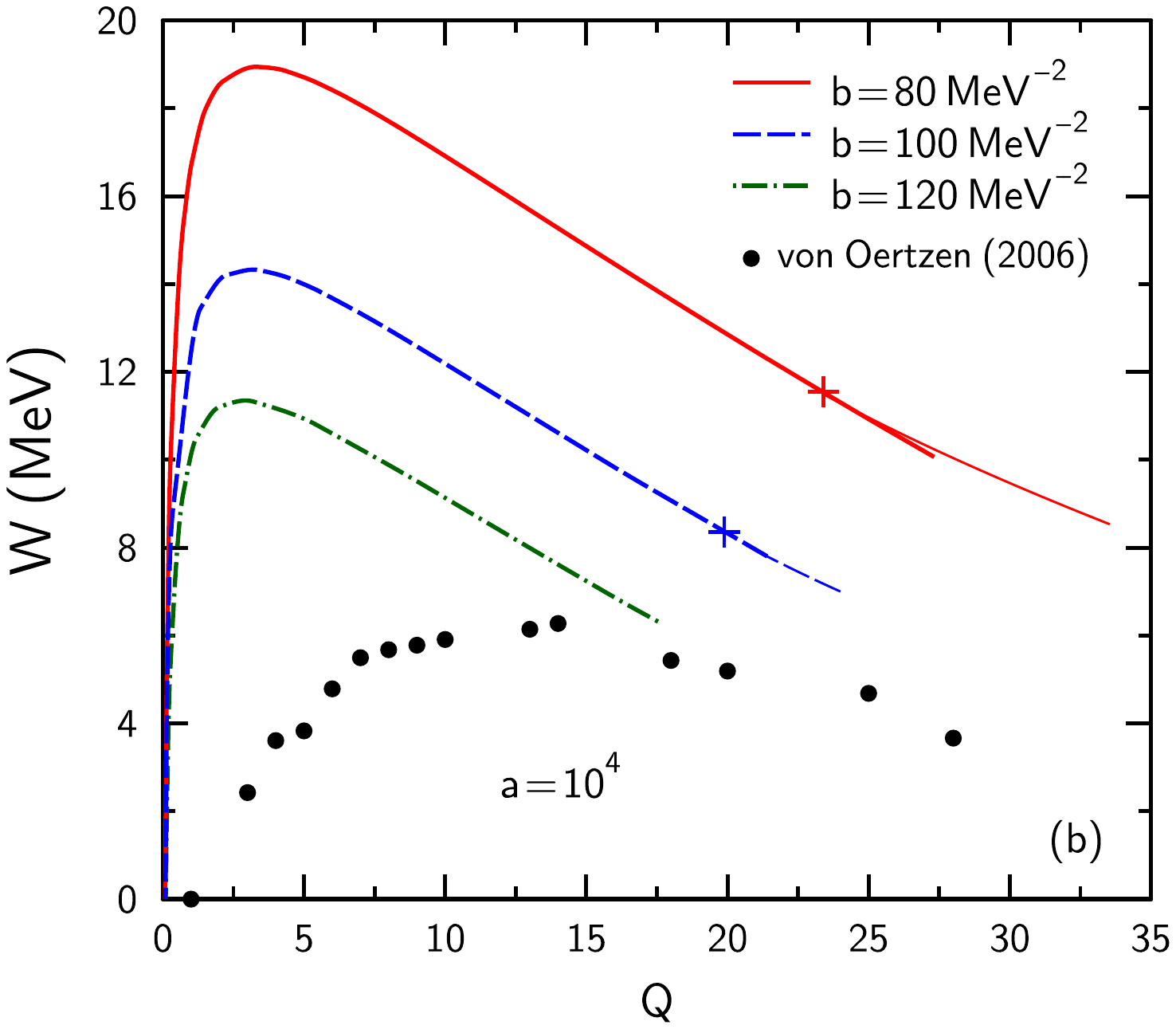}
\caption{
Binding energies per particle as functions of particle number $Q$ for different values of the parameters $a$ and $b$, {shown in the left and right panels. Thick and thin lines correspond to Q-balls and Q-shells, respectively.} Crosses mark left boundaries of domains where stable Q-shells exist.
Dots are empirical data extracted~\cite{Oer06} from observed masses of $\alpha$-conjugate nuclei.
}\label{fig6}
\end{figure}

\subsection{Modification of the gradient term}
\label{smod}

A simple {analysis} shows that the predicted surface tension of cold $\alpha$ matter, is {unrealistically} {small}, of the order of $0.2\,\textrm{MeV/fm}^2$ for the parameter Set I (see Appendix~B). This is {by a factor of 5} smaller {than} the empirical {value} for ordinary nuclei. {To avoid this drawback, we have modified}
the~GT of Lagrangian~(\ref{lagd}) {by} the replacement $(\bm{\nabla}\varphi)^2\to\xi_s(\bm{\nabla}\varphi)^2$ where
$\xi_s\geqslant 1$ is the enhancement factor. This leads to {the} appearance of additional
coefficient $\xi_s$ in front of {the} Laplacian in the KGE~(\ref{eom1}). The same factor also appears in the gradient part of {the} energy density $\varepsilon_{\rm gr}$. Our calculations show
(see {Figs.}~\ref{fig7} and \ref{fig8})
that this modification of the model leads to smoothing density profiles
and reducing binding energies {for} Q-balls with~$Q\lesssim 15$.

\begin{figure}[htb!]
\centering
\includegraphics[trim=2cm 7.5cm 3cm 8cm,width=0.48\textwidth]{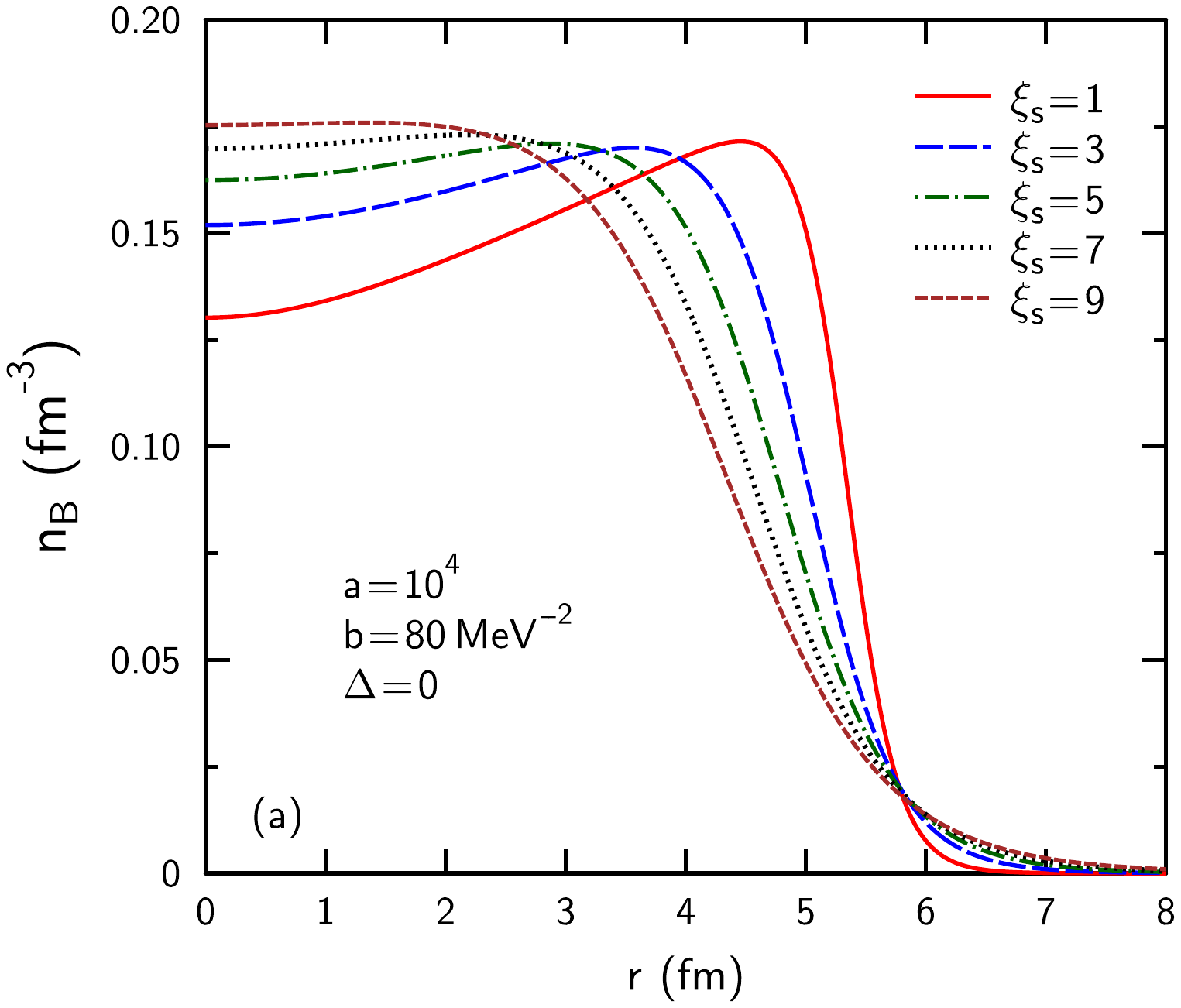}
\includegraphics[trim=2cm 7.5cm 3cm 8cm,width=0.48\textwidth]{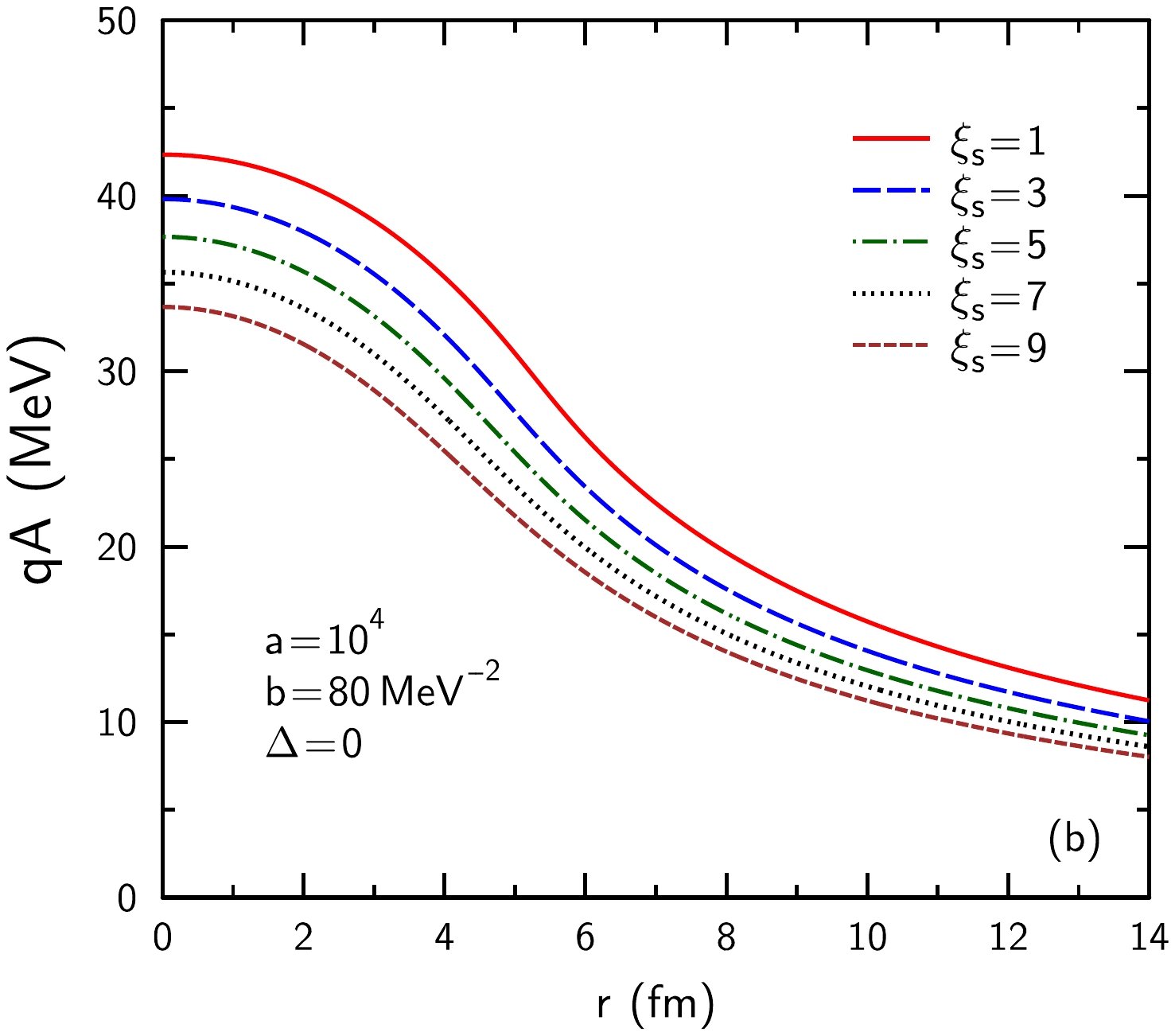}
\caption{
Profiles of the baryon density (a) and Coulomb potential (b) for different values of the gra\-dient enhancement factor~$\xi_s$.
}\label{fig7}
\end{figure}
Figure~\ref{fig8} shows that the empirical data {can be}
well reproduced with large~\mbox{$\xi_s\sim 8-9$}\,.
\begin{figure}[ht!]
\centering
\includegraphics[trim=2cm 7.5cm 3cm 8cm,width=0.6\textwidth]{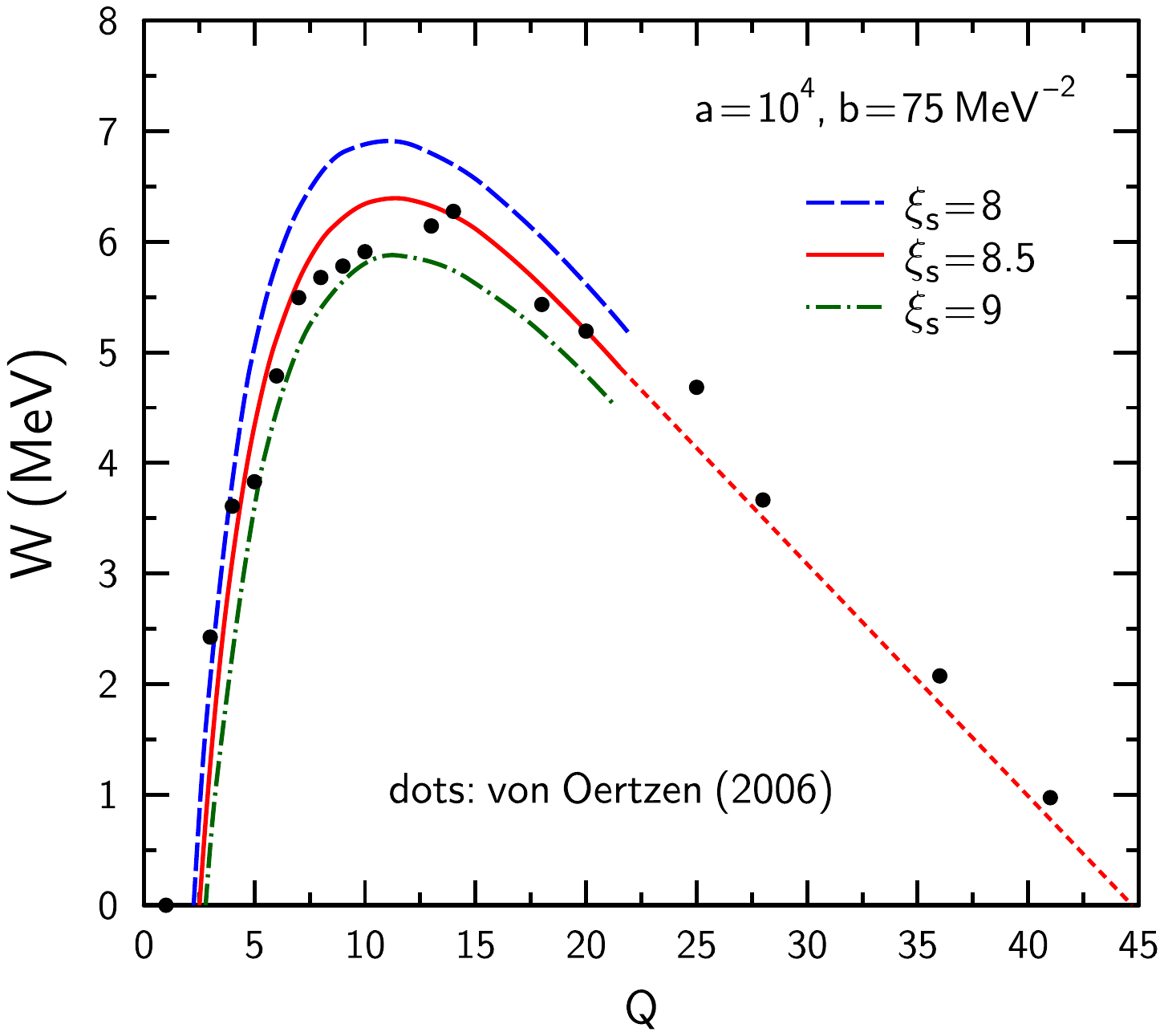}
\caption{
Binding energy $W$ as the function of $Q$ for $a=10^4,\,b=75~\textrm{MeV}^{-2}$  and $\xi_s=8,8.5$
and~$9$\hspm.
Short-dashed line is obtained by linear extrapolation of the solid curve to the region $Q>Q_{\rm max}$ where $Q_{\rm max}\simeq 21.5$ is the particle number corresponding to $\mu=m$ for $\xi_s=8.5$\,.
}\label{fig8}
\end{figure}
As~shown in Appendix~B, the resulting surface tension {coefficient}
becomes similar to that for ordinary nuclei. In Fig.~\ref{fig8} the results for {$\xi_s=8.5$ are extrapolated into the region of~metastable nuclei with $Q>Q_{\rm max}$ (the short-dashed line). Note
that the extrapolation is close to the empirical data {even} at large $Q$\,.
{From this analysis we conclude that these data can be well described by
the model parameters $a=10^4$, $b=75~\textrm{MeV}^{-2}$~\mbox{and $\xi_s=8.5$ (Set~II)}.}

\subsection{Importance of finite-size effects}
\label{smod1}

Of course, such a strong modification of the GT needs to be explained. From the density profiles in Fig.~\ref{fig7} one can see that at $\xi_s\lesssim 3$, surface widths
of Q-balls are smaller than the rms {radius of~$\alpha$~particle~$R_\alpha\simeq~1.7~\textrm{fm}$~\cite{Kra21}.} This unrealistic behavior follows from the point-like
character of strong interaction assumed in the Skyrme potential~(\ref{skpt}).
In reality, $\alpha$~particles are extended objects which can be described by a (normalized)
Gaussian density distribution
\bel{ffe}
F(r)=\dfrac{\lambda^3}{\pi^{3/2}}\exp\hsp (-\lambda^2 r^2)\,,
\ee
where the parameter $\lambda\sim R_\alpha^{-1}$. Therefore, the attractive interaction
of $\alpha$ particles is only possible at distances $\Delta r\gtrsim 2\hspm R_\alpha$\,\hspm\footnote
{
{At smaller distances, a repulsive interaction of the van der Waals type should be also included.}
}.

One can estimate this finite-size effect by adding the correction term
to the Lagrangian\hspm\footnote
{{In the limit of a point-like interaction ($\lambda\to\infty$)} one has
$F(R)\to\delta\hspace*{.2pt}(\bm{R})$  and $\delta\hspace*{.2pt}{\cal L}\to 0$\,.
}
\bel{dlch}
\delta{\cal L}=\dfrac{a}{4}\int F(R)\left[\varphi^2\left(\bm{r}+
\frac{\bm{R}}{2}\right)\hspm\varphi^2\left(\bm{r}-\frac{\bm{R}}{2}\right)
-\varphi^4(\bm{r})\right]d^{\,3}R\,,
\ee
where $R$ is the distance between the $\alpha$-particle centers.
Expanding the expression in square brackets up to the~lowest order in $R$ and neglecting the curvature terms, we obtain
\bel{dlch1}
\delta{\cal L}\simeq -\frac{a}{24}\,\varphi^2(\bm{r})\,(\bm{\nabla}\varphi)^2\int F(R)\,R^2d^{\,3}R\,.
\ee

{After adding $\delta{\cal L}$ to the Lagrangian (\ref{lagd})} one can see that the {GT is enhanced by the factor}
\bel{enh}
\xi_s\simeq 1+\frac{a}{8\hsp\lambda^2}\,\varphi^2(\bm{r})\sim 1+\dfrac{k_0^2}{4\hsp\lambda^2}\,.
\ee
Here we have taken $\varphi^2\sim\varphi_0^2/2$, where
$\varphi_0$ is defined in \re{mpar},  and introduced the charac\-teristic momentum
$k_0$ from~\re{msca}.
For the parameter set {with} \mbox{$a=10^4$}, \mbox{$b=75~\textrm{MeV}^{-2}$}, one has $k_0=2.53~\textrm{fm}^{-1}$. {Substituting $\lambda=0.5~\textrm{fm}^{-1}$
we obtain the enhancement factor~$\xi_s\sim 7.4$} which is close to the value extracted from the fit of data on $\alpha$-conjugate nuclei (see Fig.~\ref{fig8}).

\section{Estimating life times of metastable Q-balls}
\label{msqb1}

{Life} times of metastable {Q-balls} can be estimated by using standard {formalism}
for \mbox{$\alpha$-decay} of ordinary nuclei~\textbf{\cite{Gam28,Pov15}}. We
{approximate the $\alpha$-particle potential in the Q-ball by}
a~combination of an attractive square well with radius $R$ and the {repulsive} Coulomb {barrier}
\mbox{$V(r)=q^2Q/r$} at~$r>R$, where $Q$ is total number of $\alpha$ particles {in the} Q-ball.
One can estimate the single-particle (nonrelativistic) energy of $\alpha$'s in a metastable state as
\bel{enms}
E\simeq\mu-m\simeq \frac{3\hsp q^2}{2\hspm R}\hsp (Q-Q_m)>0\,,
\ee
where $Q_m$ is the maximal value of $Q$ for stable states. We neglect
the energy spreading due to {a} nonzero decay width. {It is also assumed}
that $\mu$ changes linearly with the Coulomb potential at $r=0$ (see footnote
on page~\pageref{elst1}). In the following we neglect the $Q$-dependence of $R$
estimating it {by the value} at~$Q=Q_m$\,. The condition \mbox{$0<E<V(R)$}
is fulfilled if~$Q_m<Q<3\hspm Q_m$\hspm .

In the {semiclassical} approximation,
the probability of $\alpha$ particle {with the energy $E$} to penetrate the Coulomb~barrier is
\bel{tpro}
P(E)=\exp\left\{-\frac{2}{\hbar}\int\limits_{R}^{r_{\textrm{max}}}
\sqrt{2\hspm m\left[V(r)-E\right]}\hsp dr\right\},
\ee
where $r=r_{\textrm{max}}$ is determined from the equation $V(r)=E$.
\begin{figure}[ht!]
\centering
\includegraphics[trim=2cm 7.5cm 3cm 8cm,width=0.6\textwidth]{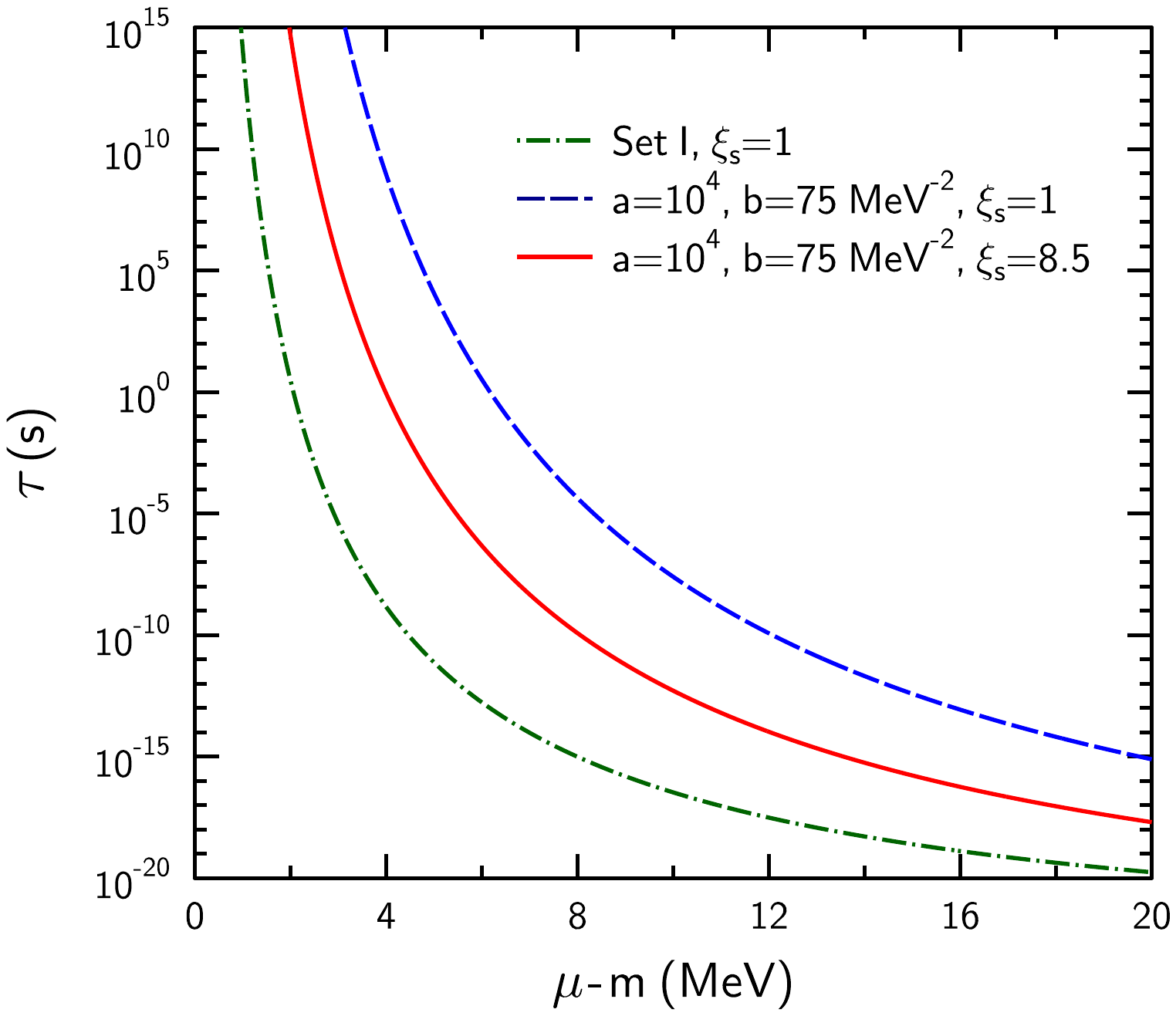}
\caption{
{Lifetimes of metastable Q-balls at $\mu>m$ for different sets of model parameters.}
}\label{fig9}
\end{figure}
The integral in \re{tpro} can be calculated analytically {that} gives
\bel{tpro1}
|\ln{P}|=\frac{2\hspm m\hspm v}{\hbar R}\,I(\alpha),~~\alpha=\frac{2}{3}\frac{Q}{Q-Q_m}\,,
\ee
where $v=\sqrt{2\hspm E/m}$ {is the mean velocity of $\alpha$ particles at $r<R$} and
\bel{int1}
I(\alpha)=\int\limits_1^\alpha\sqrt{\frac{\alpha}{x}-1}\,dx=\alpha\hsp\tan^{-1}{\sqrt{\alpha-1}}-\sqrt{\alpha-1}\,.
\ee
The life time of the metastable state with energy $E$ can be estimated as
\bel{dtim}
\tau(E)\simeq\frac{2R}{v\hsp P\hsp (E)}\,.
\ee
The results of calculating $\tau$ as a function of $\mu$ is shown in Fig.~\ref{fig9} for
different sets of model parameters.
Obviously, Q-balls closer to the threshold $\mu=m$ have longer life times,
and~\mbox{$\tau\to\infty$} at $\mu\to m$\,.
{With} the 'best fit' {parameters} corresponding to the solid line in Fig.~\ref{fig9},
we predict life times $\tau\gtrsim 10^{-10}~s$ {for} $\mu-m\lesssim 8~\textrm{MeV}$\,.
This is much longer than typical time scales in ordinary nuclei.
Such metastable {Q-balls} emitting multiple $\alpha$ particles could be good candidates for
$\alpha$-clustered nuclei.

\section{Conclusions}

In the present paper we {have formulated} the mean-field model for describing
finite{-size} systems of charged scalar bosons with Skyrme{-like effective}
interactions. {This model is used to study the Bose-Einstein condensation of
$\alpha$ particles in {drops} of nuclear size containing up~to~$50\,\alpha$ clusters.}
{The baryon} density, energy density, and effective mass profiles {{were} calculated for}
different values of the chemical potential $\mu$. Two types of solutions have been found:
\mbox{Q-balls} with nonzero density at the center and Q-shells with vanishing density
in the central region. It is shown that {stable} Q-shells appear only for large enough
strengths of~attractive interaction.

We {have calculated} the GS {binding energies} of \mbox{Q-balls} (\mbox{Q-shells}) as functions of {$\mu$} and {investigated} their stability regions.
Both stable (with \mbox{$\mu<m$}, where~$m$~is the $\alpha$-particle mass)
and metastable ($\mu>m$) solutions were considered. It~{was} shown that life times of metastable Q-balls {may} be rather long in nuclear scale. Such Q-balls are especially interesting due to
the possibility of simultaneous emission of several $\alpha$ particles. These decay
channels could be a~unique signature of $\alpha$ clustering in nuclei.

We {have} tried to find the set of model parameters which would fit the empirical binding energies of $\alpha$-conjugate nuclei {compiled} in Ref.~\cite{Oer06}.
{It turned out that} the standard version of~the model {with}
point-like $\alpha$ particles strongly overestimates binding {energies} of lighter \mbox{Q-balls}. The agreement with empirical data has been achieved {only} when the gradient term of the Lagrangian was significantly enhanced. {It is demonstrated} that this enhancement {can} be naturally explained by {a} finite size of $\alpha$'s. {We would like to note, that our assumption that ground states of $\alpha$-conjugate nuclei} {contain only $\alpha$ particles and no} {nucleons is probably} {oversimplified}. {In the future we are
going to study more complicated} {finite-size} {alpha-nucleon systems.}

\begin{acknowledgments}
The authors thank {M.~Gorenstein, S.~Misicu, and O.~Savchuk for useful discussions.}
L.M.S., and I.N.M.~appreciates the support from the Frankfurt Institute for Advanced Studies.
H.St.~thanks the support from Walter Greiner Gesellshaft zur F\"orderung der physikalishen
Grundlagen e.V. through the J.~M.~Eisenberg Laureatus chair at Goethe Universit\"at Frankfurt am Main.
\end{acknowledgments}

\section*{Appendix~A: Numerical procedure}
\setcounter{equation}{0}
\renewcommand{\theequation}{A\arabic{equation}}

In this section we describe our numerical procedure {for calculating} properties of {(meta)}sta\-ble
\mbox{Q-balls} ({and/or}~Q-shells). {For such calculations it} is important to impose suitable asymptotic conditions at large~$r$. At $r\to\infty$ one can write approximately
$U\simeq (m\varphi)^2/2$\hspm , and $A\simeq q\hsp Q/r$, where $Q$ is the total number of $\alpha$ particles
bound in {the} Q-ball. Then using \re{eom1}, one finds following asymptotic {behaviour} (see for details Ref.~\cite{Gul15})
\bel{ascn}
\varphi\simeq\textrm{const}\,\dfrac{e^{-\sqrt{m^2-\mu^2}\hspm r}}{r^{1+\beta\hspm Q}},~~
\beta=\frac{q^2\mu}{\sqrt{m^2-\mu^2}}~~~(\textrm{at}~r\to\infty)\hsp .
\ee
One can see that at \mbox{$\mu>m$} the scalar field becomes a complex function which results in a~finite flux at \mbox{$r\to\infty$}. {However,} at not too large \mbox{$\mu-m$} one can consider
such systems as metastable nuclei {emitting} $\alpha$'s from their surface. Due to presence of the Coulomb barrier, corresponding decay times may be rather long, {as demonstrated in Sec.~\ref{msqb1}}.

It is useful to define a new function $Q_*(r)$ which is equal to the number of particles
within a sphere of radius $r$ (note that $Q_*(0)=0$ and \mbox{$Q_*\to Q$} at $r\to\infty$)\,. According to the Gauss law
of electrostatics, $A'(r)=-q\hsp Q_*(r)/r^2$\,.
One can rewrite Eqs.~(\ref{eom1})--(\ref{eom2}) by introducing the dimensionless quantities
$\rho=k_0\hspm r, f=\varphi/\varphi_0, g=(\mu-q\hspm A)/k_0$\,, where
\bel{msca}
k_0=\sqrt{m^2-\mu_0^2}=\frac{m}{\sqrt{\Lambda}},
\ee
is the characteristic momentum scale of the equilibrium $\alpha$ matter\hspm\footnote
{
Here we denote its chemical potential as $\mu_0$ and apply~\re{inma1} in the second equality.
}.
We arrive at the following set of differential equations
which determine the profiles of $f,g$, and~$Q_*$
\begin{eqnarray}
&&\frac{d^2f}{d\rho^2}+\frac{2}{\rho}\frac{df}{d\rho}=f\left(\Lambda-g^2-4f^2+3f^4\right),
~~\label{feq1}\\
&&\frac{d\hspm g}{d\rho}=\dfrac{q^2Q_*}{\rho^2}\,,~~~~\label{geq1}\\
&&\frac{d\hspm Q_*}{d\rho}=\dfrac{16\pi}{a}\,g\,(f\rho)^2.~~~~\label{qeq1}
\end{eqnarray}

By introducing a spatial grid $\rho_i=h\hsp i\,,~i=0,1,\ldots i_{\rm max}$, where
$h\sim 10^{-2}$ and \mbox{$i_{\rm max}\sim 5\cdot 10^3$},
we {approximated
Eqs.~(\ref{feq1})--(\ref{qeq1}) by {an} algebraic system of finite-difference equations.
{It}~was solved by iterations, using the program package 'dsolve' from Numerical Recipes~\cite{Tef98}.
The initial profiles of $f\hspm (\rho)$ and $g\hsp (\rho)$ were {chosen} by using semi-analytic
approximations, suggested in Refs.~\cite{Hee21a,Hee21b}.

\section*{Appendix~B: Surface tension of Q-balls}
\setcounter{equation}{0}
\renewcommand{\theequation}{B\arabic{equation}}

In this section we analyze the surface tension $\sigma_s$ of large Q-balls neglecting Coulomb
field in the KGE (\ref{eom1}). In fact, we {derive} the surface
tension of a cold condensate of~(uncharged) bosons interacting with the Skyrme potential (\ref{skpt}).
We {calculate} $\sigma_s$ {as follows}
\bel{stcd}
\sigma_s=(4\pi R^2)^{-1}\int\varepsilon_{\rm gr}\,d^3r
\simeq\int\limits_0^{\infty}\left(\dfrac{d\varphi}{dz}\right)^2 dz\,,
\ee
where $\varepsilon_{\rm gr}$ is defined in~\re{endc}. In the second equality,
instead of a spherical Q-ball with radius $R$, a one-dimensional symmetrical slab
\mbox{$-\infty<z<+\infty$} {is considered}. Such an~appro\-ximation is justified {for} large $Q$.
{In this case the} scalar field $\varphi\hspm (z)$ obeys the one-dimensional~KGE
\bel{kge2d}
\frac{d^2\varphi}{dz^2}+\mu^2\varphi=\frac{dU}{d\varphi}\,.
\ee
The first integral of this equation is easily obtained
\bel{kges2}
\frac{1}{2}\left(\frac{d\varphi}{dz}\right)^2=U\hsp (\varphi)-\frac{\mu^2\varphi^2}{2}
\equiv\Omega\hsp (\varphi)\hsp\,.
\ee
Here we have taken into account that $\varphi,d\varphi/dz\to 0$ at $|z|\to\infty$\,. In fact, this
equation implies that pressure $p=T_{zz}=0$ {for} all $z$\hspm . In the last equality
of (\ref{kges2}) we introduce the thermodynamic potential $\Omega=\varepsilon-\mu\hspm n$
for a~homogeneous scalar field $\varphi$ (see Sec.~\ref{infm}). It is easy to see that
at the central plane, $z=0$, one has $d\varphi/dz=0$ and $\varphi=\varphi_*$\hsp, where~$\varphi_*$
is found from the equation $\Omega\hsp (\varphi_*)=0$\hsp .

Finally {one obtains} the relation (first derived in Ref.~\cite{Col85})
\bel{stcd1}
\sigma_s=\int\limits_0^{\varphi_*}\sqrt{2\hsp\Omega (\varphi)}\,d\varphi\,.
\ee
In the case of Skyrme interaction, substituting (\ref{skpt}), one has
\bel{stcd2}
\sigma_s=k_0\varphi_0^2\int\limits_0^{f_*}\sqrt{\eta^2-2f^2+f^4}fdf\,,
\ee
{where}
\bel{not1}
f_*=\sqrt{1-\sqrt{1-\eta^2}},~~~\eta=\frac{1}{k_0}\sqrt{m^2-\mu^2}\,.
\ee
Here we use the constants $\varphi_0, k_0$ defined in Eqs.~(\ref{mpar}), (\ref{msca}).

The integral in \re{stcd2} can be calculated analytically. One gets
\bel{stcd3}
\sigma_s=\sigma_0\hsp F(\eta),
\ee
where
\bel{not2}
\sigma_0=\frac{k_0\varphi_0^2}{4}=a^2\left(\frac{3}{16\hsp b}\right)^{3/2},
\ee
and
\bel{funf}
F(\eta)=\eta-(1-\eta^2)\ln\hsp\sqrt{\frac{1+\eta}{1-\eta}}
\ee
is a dimensionless function which monotonically increases from zero to unity
in the interval~\mbox{$0<\eta<1$}. Equations~(\ref{stcd3})--(\ref{funf}) give the analytic expression
for $\sigma_s$ as a function of chemical potential. It is interesting to note that heaviest
stable Q-balls with $\mu=m$ have vanishing surface tension. One can see that $\sigma_s$ reaches its maximal value $\sigma_0$ for equilibrium bosonic matter with $\mu=\mu_0$\,.

Numerical estimates give rather small values for $\sigma_0$\hspm . For example, by choosing the
parameters~$a,b$ from Set I, one obtains \mbox{$\sigma_0\simeq 0.18~\textrm{MeV/fm}^2$}. A somewhat larger value, \mbox{$\sigma_0\simeq 0.32~\textrm{MeV/fm}^2$}, is obtained for $a=10^4, b=75~\textrm{MeV}^{-2}$.
On the other hand, much higher surface tension coefficients are expected for cold ordinary nuclei. {Indeed, \mbox{according}} to the Weizs\"acker mass formula, the nuclear surface energy
equals $E_S=a_S\hsp A^{2/3}$ where $A$ is the baryon number of the nucleus and
$a_S\simeq 17.2~\textrm{MeV}$~\cite{Mar19}. Dividing this energy by $4\pi R^2$
(here $R=r_0\hsp A^{1/3}\simeq 1.2 \hsp A^{1/3}~\textrm{fm}$ is the geometrical radius of the nucleus), one {obtains} the nuclear surface tension coefficient
\bel{stnu}
\sigma_{sN}\simeq \frac{a_S}{4\pi r_0^2}\simeq 0.95~\textrm{MeV/fm}^2.
\ee
This exceeds the {value} of $\sigma_s$ {predicted for Q-balls} by about a factor of three.

Such a discrepancy can be removed by modifying the Laplacian term in the KGE.
One can {see} that introducing the enhancement factor $\xi_s$
{in front of Laplacian in~\re{eom1}} {leads}
to the additional coefficient $\sqrt{\xi_s}$ in the right hand sides of~Eqs.~({\ref{stcd1}})~{and}~({\ref{not2}}). Using the value $\xi_s\simeq 8.5$, which enables a good agreement with von Oertzen data (see Fig.~\ref{fig8}), we get the estimate
\mbox{$\sigma_0\simeq 0.94~\textrm{MeV/fm}^2$}, very close to that obtained in~\re{stnu}.


\end{document}